%% file: merged.tex
\documentclass[%
 10pt,
 superscriptaddress,
 reprint,
 bibnotes,
 amsfonts,
 amssymb,
 amsmath,
 aps,
 prl,
 floatfix,
]{revtex4-2}
\usepackage{graphicx}
\usepackage{xcolor}
\usepackage{tikz}
\usetikzlibrary{quantikz2}
\usepackage[
 colorlinks=true,%
 citecolor=cyan,%
 urlcolor=cyan,%
 linkcolor=cyan,%
]{hyperref}
\usepackage{array}
\usepackage{dcolumn}
\usepackage{multirow}
\usepackage{booktabs}
\usepackage{enumerate}
\usepackage[inline]{enumitem}
\usepackage{csquotes}
\usepackage{bm}
\usepackage{url}

\usepackage[normalem]{ulem}

\newcounter{one}
\newcommand{\ketbra}[2]{| {#1} \rangle\langle {#2} |}

\newcommand{\ti}[1]{\textit{#1}}
\newcommand{\tb}[1]{\textbf{#1}}

\newcommand{\ra}{\rightarrow}

\newcommand{\Tr}{\mathrm{Tr}}

\newcommand{\nd}{\textendash}

\def\QED{\mbox{\rule[0pt]{1.5ex}{1.5ex}}}

\def\endproof{\hspace*{\fill}~\QED\par\endtrivlist\unskip}

\newcommand{\calD}{\mathcal{D}}

\newcommand{\one}{I}
\newcommand{\two}{I\hspace{-0.3pt}I}
\newcommand{\three}{I\hspace{-0.3pt}I\hspace{-0.3pt}I}

\newenvironment{myquote}[1]%
  {\list{}{\leftmargin=#1\rightmargin=#1}\item[]}%
  {\endlist}

\definecolor{lightblue}{rgb}{0.678, 0.847, 0.902}
\definecolor{lightgreen}{rgb}{0.565, 0.933, 0.565}
\definecolor{lightyellow}{rgb}{1.000, 1.000, 0.600}
\definecolor{lightpurple}{rgb}{0.867, 0.627, 0.867}
\definecolor{lightorange}{rgb}{1.000, 0.753, 0.502}
\definecolor{lightpink}{rgb}{1.000, 0.714, 0.757}
\definecolor{lightred}{rgb}{1.000, 0.714, 0.757}
\definecolor{lightcyan}{rgb}{0.878, 1.000, 1.000}
\definecolor{lightorange}{rgb}{1.000, 0.753, 0.502}
\definecolor{lightcoral}{rgb}{0.941, 0.502, 0.502}
\definecolor{lightsalmon}{rgb}{1.000, 0.627, 0.478}

\newsavebox{\boxA}
\newsavebox{\boxB}
\newsavebox{\boxC}
\newsavebox{\boxD}
\newsavebox{\boxE}
\newsavebox{\boxF}
\newsavebox{\boxG}

\begin{document}
\include{main}
\clearpage
\onecolumngrid
\maketitle
\include{supp}
\end{document}

%% file: main.tex
\title{Disturbance Evaluation Circuit in Quantum Measurement}
\affiliation{
Graduate School of Information Science and Technology, Hokkaido University, Kita-ku, Sapporo, Hokkaido 060-0814, Japan
}
\affiliation{
Center for Mathematical Science and Artificial Intelligence, Chubu University Academy of Emerging Sciences, Chubu University, Matsumoto-cho, Kasugai, Aichi 487-8501, Japan
}
\affiliation{
RIKEN Innovation Design Office, Hirosawa, Wako, Saitama 351–0198, Japan
}
\affiliation{
Graduate School of Informatics, Nagoya University, Chikusa-ku, Nagoya, Aichi 464-8601, Japan
}
\author{
Haruki Emori
}
\email{emori.haruki.i8@elms.hokudai.ac.jp}
\affiliation{
Graduate School of Information Science and Technology, Hokkaido University, Kita-ku, Sapporo, Hokkaido 060-0814, Japan
}
\affiliation{
RIKEN Innovation Design Office, Hirosawa, Wako, Saitama 351–0198, Japan
}
\author{
Masanao Ozawa
}
\email{ozawa@isc.chubu.ac.jp}
\affiliation{
Center for Mathematical Science and Artificial Intelligence, Chubu University Academy of Emerging Sciences, Chubu University, Matsumoto-cho, Kasugai, Aichi 487-8501, Japan
}
\affiliation{
RIKEN Innovation Design Office, Hirosawa, Wako, Saitama 351–0198, Japan
}
\affiliation{
Graduate School of Informatics, Nagoya University, Chikusa-ku, Nagoya, Aichi 464-8601, Japan
}
\author{
Akihisa Tomita
}
\email{tomita@ist.hokudai.ac.jp}
\affiliation{
Graduate School of Information Science and Technology, Hokkaido University, Kita-ku, Sapporo, Hokkaido 060-0814, Japan
}
\date{\today}

\begin{abstract}
%The performance evaluation of quantum measurements is essential for the advancement of quantum information processing.
%In this study, we propose a novel direct evaluation method for the operator-based quantum root-mean-square (QRMS) disturbance and compare its performance with the existing approaches, namely the three-state method (TSM) and the weak measurement method (WMM).
%Our method establishes a correspondence between the QRMS disturbance of the measurement and the second-order derivative of the decoherence induced in a newly introduced weak probe system with respect to the coupling strength of the weak interaction at its zero-limit.
%Furthermore, we demonstrate the effectiveness of our method through a simulation based on a platform for quantum computing.
%The simulation results capture the key features of the TSM, WMM, and our method, providing insights into the strengths and limitations of these methods.
According to the uncertainty principle, every quantum measurement accompanies disturbance.
In particular, accurate sequential measurements need the accurate control of disturbance.
However, the correct role of disturbance in the uncertainty principle has been known only recently.
Understanding the disturbance is crucial for understanding the fundamentals of physics, and accurately evaluating the disturbance is important for quantum technologies such as quantum information processing and quantum metrology.
Therefore, the experimental evaluation of the disturbance is a significant challenge in those fields.
In this study, we propose a novel evaluation method for the quantum root-mean-square (QRMS) disturbance and compare its performance with the existing approaches, known as the three-state method (TSM) and the weak measurement method (WMM).
Our method establishes a correspondence between the QRMS disturbance of the measurement and the second-order derivative of the decoherence induced in a newly introduced weak probe system with respect to the coupling strength of the weak interaction at its zero-limit.
Furthermore, we demonstrate the effectiveness of our method in comparison with the other two through a simulation and experiment using a quantum computer.
The results capture the key features of the TSM, WMM, and our method, providing insights into the strengths and limitations of these methods.
\end{abstract}

\maketitle

\section{\label{sec:introduction}Introduction}
Originating from Heisenberg's seminal paper \cite{Heisenberg27}, the experimentally testable formulation of the uncertainty principle capturing the essence of incompatible quantum measurements has been developed with the quantum root-mean-square (QRMS) error and disturbance based on error and disturbance operators \cite{Ozawa02,Ozawa03,Ozawa04,Branciard13}.

The operator-based QRMS error of the observable to be measured is defined as the square-root of the second moment of the error operator -- the difference between the post-measurement Heisenberg operator of the meter observable and the pre-measurement Heisenberg operator of the observable to be measured (the measured observable).

The operator-based QRMS disturbance of an observable is defined as the square-root of the disturbance operator -- the difference between the post-measurement Heisenberg operator and the pre-measurement Heisenberg operator of the observable to be disturbed (the disturbed observable).

In particular, the operator-based QRMS error is sound and complete when the post-measurement meter observable and the pre-measurement measured observable commute.
On the other hand, it is sound but not complete when they do not commute \cite{Ozawa19NPJ}.
Similarly, the operator-based QRMS disturbance is sound and complete when the post-measurement disturbed observable and the pre-measurement disturbed observable commute.
On the other hand, it is sound but not complete when they do not commute \cite{Ozawa21}.
To obtain sound and complete QRMS (error and disturbance) measures, we have introduced locally uniform QRMS measures \cite{Ozawa19NPJ,Ozawa21}, which maximise the operator-based QRMS measures `locally', i.e. over the orbit generated by the observable and the state to be measured.
The above maximisations have been experimentally realised by sweeping the input state for the evaluations of the operator-based QRMS measures \cite{Sponar21}.
Therefore, the operator-based QRMS measures are indispensable notions not only in the commutative case but also in the noncommutative case as the basis for locally uniform QRMS measures \cite{Ozawa19NPJ,Ozawa21,Ozawa19ARX}.

To experimentally evaluate the QRMS measures, two methods have been employed: the \ti{three-state method} (TSM) \cite{Ozawa04} and the \ti{weak measurement method} (WMM) \cite{Ozawa05PLA,Lund10}, which are depicted in Figs.~\ref{fig:dem}(a) and (b), respectively.
In the TSM, we prepare three states for the first apparatus, and then, for the corresponding output states, we measure the first and second moments of the outputs from the second apparatus \cite{Erhart12,Sulyok13,Baek13,Demirel16}.
In the WMM, we perform the weak and strong measurements before and after the first apparatus and calculate the weak joint distribution (WJD) \cite{Rozema12,Weston13,Kaneda14,Ringbauer14}.

In addition to the above two methods for evaluating the QRMS measures, for the QRMS error, a direct evaluation method has been devised recently \cite{Hofmann21,Lemmel22} through the introduction of a feedback operation establishing a correspondence between the error of the measurement and the second-order derivative of the decoherence induced in a newly introduced weak probe system, with respect to the coupling strength of the weak interaction at its zero-limit -- hereafter, we refer to it as the second-order coefficient of the decoherence.
In contrast, for the QRMS disturbance, its counterpart, the corresponding direct evaluation method, is not yet known.

In this study, we propose a novel direct evaluation method for the QRMS disturbance.
We use the weak interaction represented by the unitary operator (generated by the tensor product of the \ti{disturbed observable} and the Pauli-Z operator) acting on the \ti{measured system} plus the \ti{weak probe system} after the measuring interaction, without the feedback operation based on measurement results.

Our method establishes a correspondence between the \ti{disturbance} of the measurement and the \ti{second-order coefficient of the decoherence}.
It is noteworthy that, in addition to the TSM and WMM, our method naturally becomes effective for the evaluation of the locally uniform QRMS disturbance through the sweeping technique mentioned above (see Supplementary Information Sec.~\one~A).

In addition to the theoretical proposal, we present the numerical results of our method, along with those of the TSM and WMM, obtained using the superconducting quantum computer \texttt{ibm\_kawasaki} of IBM Q \cite{IBMQ}.
While the TSM and WMM have been used for experimental investigations of uncertainty relations \cite{Erhart12,Sulyok13,Baek13,Demirel16,Rozema12,Weston13,Kaneda14,Ringbauer14}, the statistical properties of the data obtained by those methods have not been explored.
By analysing the numerical results, we can understand the performance of the TSM, WMM, and our method in comparison.
Moreover, this enables us to discuss the compatibility between quantum error mitigation techniques and the corresponding evaluation methods for errors and disturbances.
The comparison data will be useful for benchmarking quantum information processing devices in the future.

\savebox{\boxA}{
\resizebox{0.9\linewidth}{!}{
\begin{quantikz}[thin lines] 
\lstick{\textbf{(a)}\ $\sigma_{\mathbf{P}}$} & \qw & \qw & \qw & \gate[2]{U} \gategroup[2, steps = 2, style = {dashed, rounded corners, fill = lightpurple, inner xsep = 2pt, inner ysep = 2pt}, background, label style = {label position = below, anchor = north, yshift = -0.2cm}]{{\footnotesize{Indirect Meas.}}} & \meterD{M} & \qw & \qw \\
\lstick{$\rho_{\mathbf{S}}$} & \gate{\{I,B,I+B\}} \gategroup[1, steps = 1, style = {dashed, rounded corners, fill = lightgreen, inner xsep = 2pt, inner ysep = 2pt}, background, label style = {label position = below, anchor = north, yshift = -0.2cm}]{{\footnotesize{Three-state gen. op.}}} & \qw & \qw & \qw & \qw & \qw & \meterD{B} \gategroup[1, steps = 1, style = {dashed, rounded corners, fill = lightyellow, inner xsep = 2pt, inner ysep = 2pt}, background, label style = {label position = below, anchor = north, yshift = -0.2cm}]{{\footnotesize{Ideal Meas.}}}
\end{quantikz}
}}

\savebox{\boxB}{
\resizebox{0.9\linewidth}{!}{
\begin{quantikz}[thin lines] 
\lstick{\textbf{(b)}\ $\sigma_{\mathbf{P}}$} & \qw & \qw & \qw & \gate[2]{U} \gategroup[2, steps = 2, style = {dashed, rounded corners, fill = lightpurple, inner xsep = 2pt, inner ysep = 2pt}, background, label style = {label position = below, anchor = north, yshift = -0.2cm}]{{\footnotesize{Indirect Meas.}}} & \meterD{M} & \qw & \qw \\
\lstick{$\rho_{\mathbf{S}}$} & \gate[2]{W(\theta_{\text{w}})} \gategroup[2, steps = 2, style = {dashed, rounded corners, fill = lightblue, inner xsep = 2pt, inner ysep = 2pt}, background, label style = {label position = below, anchor = north, yshift = -0.2cm}]{{\footnotesize{Weak Meas.}}} & \qw & \qw & \qw & \qw & \qw & \meterD{B} \gategroup[1, steps = 1, style = {dashed, rounded corners, fill = lightyellow, inner xsep = 2pt, inner ysep = 2pt}, background, label style = {label position = below, anchor = north, yshift = -0.2cm}]{{\footnotesize{Ideal Meas.}}} \\
\lstick{$\xi_{\mathbf{P'}}$}& \qw & \meterD{M'} & \qw & \qw & \qw & \qw & \qw 
\end{quantikz}
}}

\savebox{\boxC}{
\resizebox{\linewidth}{!}{
\begin{quantikz}[thin lines] 
\lstick{\textbf{(c)}\ $\sigma_{\mathbf{P}}$} & \qw & \qw & \gate[2]{U} \gategroup[2, steps = 2, style = {dashed, rounded corners, fill = lightpurple, inner xsep = 2pt, inner ysep = 2pt}, background, label style = {label position = below, anchor = north, yshift = -0.2cm}]{{\footnotesize{Indirect Meas.}}} & \meterD{M} & \qw & \qw & \qw & \qw \\
\lstick{$\rho_{\mathbf{S}}$} & \gate[2]{V(\theta)} \gategroup[2, steps = 1, style = {dashed, rounded corners, fill = lightpink, inner xsep = 2pt, inner ysep = 2pt}, background, label style = {label position = below, anchor = north, yshift = -0.2cm}]{{\footnotesize{Weak Int.}}} & \qw & \qw & \qw & \qw & \gate[2]{V^{\dagger}(\theta)} \gategroup[2, steps = 1, style = {dashed, rounded corners, fill = lightpink, inner xsep = 2pt, inner ysep = 2pt}, background, label style = {label position = below, anchor = north, yshift = -0.2cm}]{{\footnotesize{Weak Int.}}} & \qw & \qw \\
\lstick{$|+\rangle_{\mathbf{P'}}$} & \qw & \qw & \qw & \qw & \qw & \qw & \qw & \meterD{P^{X}_{+}} \gategroup[1, steps = 1, style = {dashed, rounded corners, fill = lightorange, inner xsep = 2pt, inner ysep = 2pt}, background, label style = {label position = below, anchor = north, yshift = -0.2cm}]{{\footnotesize{Ideal Meas.}}}
\end{quantikz}
}}

\begin{figure}[tb]
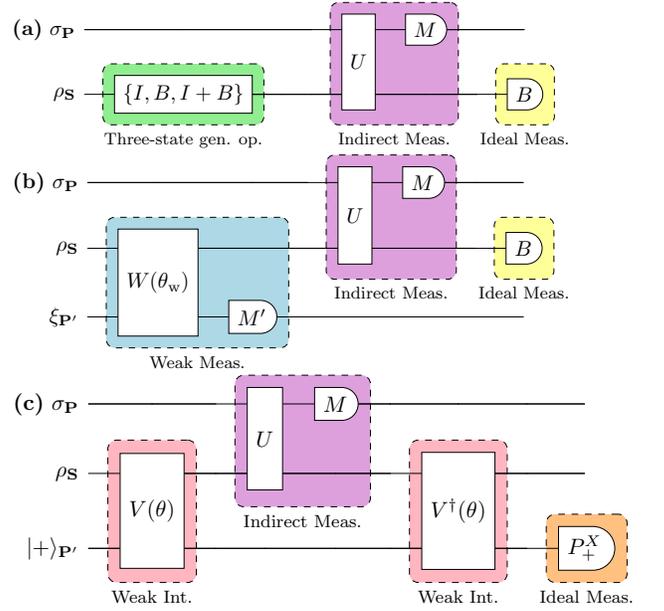

\centering
\begin{tabular}{l}
\usebox\boxA \\
\usebox\boxB \\
\usebox\boxC
\end{tabular}
\caption{
Schematic diagrams of the TSM, WMM, and DEC.
\textbf{(a)} TSM: Apply one of the three operations $\{I, B, I+B\}$ to $\mathbf{S}$ for the preparation of three states and perform expectation value sampling after the measurement $\{M_{m}\}$. 
\textbf{(b)} WMM: Apply the weak measurements, the measurement $\{M_{m}\}$, and the strong measurement to $\mathbf{S}$ and compute the weak joint distribution.
\textbf{(c)} DEC: Apply the weak interaction $V(\theta)$ to $\mathbf{S}+\mathbf{P'}$, the measurement $\{M_{m}\}$ to $\mathbf{S}$, and the inverse weak interaction $V^{\dagger}(\theta)$ to $\mathbf{S}+\mathbf{P'}$ and obtain the expectation value $\langle P^{X}_{+} \rangle^\text{Out}_{\mathbf{P'}}$.
}\label{fig:dem}
\end{figure}

\savebox{\boxD}{
\resizebox{\linewidth}{!}{
\begin{quantikz}[thin lines]
\lstick{\textbf{(a)}} & \qw & \qw & \qw & \qw \gategroup[2, steps = 3, style = {dashed, rounded corners, fill = lightpurple, inner xsep = 2pt, inner ysep = 2pt}, background, label style = {label position = below, anchor = north, yshift = -0.2cm}]{} & \targ{} & \meter{}\\
& \gate{\mathsf{S^{\dagger}}} & \gate{\mathsf{H}} & \gate{\text{TSGO}} \gategroup[1, steps = 1, style = {dashed, rounded corners, fill = lightgreen, inner xsep = 2pt, inner ysep = 2pt}, background, label style = {label position = below, anchor = north, yshift = -0.2cm}]{} & \gate{\mathsf{H}} & \ctrl{-1} & \gate{\mathsf{H}} & \gate{\mathsf{H}} \gategroup[1, steps = 2, style = {dashed, rounded corners, fill = lightyellow, inner xsep = 2pt, inner ysep = 2pt}, background, label style = {label position = below, anchor = north, yshift = -0.2cm}]{} & \meter{}
\end{quantikz}$\Biggl($
\begin{quantikz}[thin lines]
& \gate{\text{TSGO}} &
\end{quantikz}=\begin{quantikz}[thin lines]
& \qw,
\end{quantikz}\begin{quantikz}[thin lines]
& \gate{X} &,
\end{quantikz}\begin{quantikz}[thin lines]
& \qw & \targ{} & \meter{} \rstick{=0} \\
& \gate{\mathsf{H}} & \ctrl{-1} & \gate{\mathsf{H}} & \qw
\end{quantikz}$\Biggr)$
}}

\savebox{\boxE}{
\resizebox{0.65\linewidth}{!}{
\begin{quantikz}[thin lines, wire types={q,q,n}]
\lstick{\textbf{(b)}} & \qw & \qw & \qw & \qw & \qw & \qw & \qw \gategroup[2, steps = 3, style = {dashed, rounded corners, fill = lightpurple, inner xsep = 2pt, inner ysep = 2pt}, background, label style = {label position = below, anchor = north, yshift = -0.2cm}]{} & \targ{} & \meter{} \\
& \gate{\mathsf{S^{\dagger}}} & \gate{\mathsf{H}} & \qw \gategroup[2, steps = 4, style = {dashed, rounded corners, fill = lightblue, inner xsep = 2pt, inner ysep = 2pt}, background, label style = {label position = below, anchor = north, yshift = -0.2cm}]{} & \gate{\mathsf{H}} & \ctrl{1} & \gate{\mathsf{H}} & \gate{\mathsf{H}} & \ctrl{-1} & \gate{\mathsf{H}} & \gate{\mathsf{H}} \gategroup[1, steps = 2, style = {dashed, rounded corners, fill = lightyellow, inner xsep = 2pt, inner ysep = 2pt}, background, label style = {label position = below, anchor = north, yshift = -0.2cm}]{} & \meter{} \\
\qw & \qw & \qw & \gate{\mathsf{R_{X}(\theta)}} \qw & \gate{\mathsf{S}} \qw & \targ{} \qw & \meter{} \qw & & & & &
\end{quantikz}
}}

\savebox{\boxF}{
\resizebox{\linewidth}{!}{
\begin{quantikz}[thin lines]
\lstick{\textbf{(c)}} & \qw & \qw & \qw & \qw & \qw & \qw & \qw & \qw \gategroup[2, steps = 3, style = {dashed, rounded corners, fill = lightpurple, inner xsep = 2pt, inner ysep = 2pt}, background, label style = {label position = below, anchor = north, yshift = -0.2cm}]{} & \targ{} & \meter{} \\
& \gate{\mathsf{S^{\dagger}}} & \gate{\mathsf{H}} & \gate{\mathsf{H}} \gategroup[2, steps = 5, style = {dashed, rounded corners, fill = lightpink, inner xsep = 2pt, inner ysep = 2pt}, background, label style = {label position = below, anchor = north, yshift = -0.2cm}]{} & \ctrl{1} & \qw & \ctrl{1} & \gate{\mathsf{H}} & \gate{\mathsf{H}} & \ctrl{-1} & \gate{\mathsf{H}} & \gate{\mathsf{H}} \gategroup[2, steps = 5, style = {dashed, rounded corners, fill = lightpink, inner xsep = 2pt, inner ysep = 2pt}, background, label style = {label position = below, anchor = north, yshift = -0.2cm}]{} & \ctrl{1} & \qw & \ctrl{1} & \gate{\mathsf{H}} & \qw & \qw \\
& \gate{\mathsf{H}} & \qw & \qw & \targ{} & \gate{\mathsf{R_{Z}(\theta)}} & \targ{} & \qw & \qw & \qw & \qw & \qw & \targ{} & \gate{\mathsf{R_{Z}(-\theta)}} & \targ{} & \qw & \gate{\mathsf{H}} \gategroup[1, steps = 2, style = {dashed, rounded corners, fill = lightorange, inner xsep = 2pt, inner ysep = 2pt}, background, label style = {label position = below, anchor = north, yshift = -0.2cm}]{} & \meter{}
\end{quantikz}
}}

\savebox{\boxG}{
\resizebox{0.85\linewidth}{!}{
\begin{quantikz}[thin lines]
\lstick{\textbf{(d)}} & \qw & \targ{} & \meter{} \\
& \gate{\mathsf{H}} & \ctrl{-1} & \gate{\mathsf{H}} & \qw
\end{quantikz}$=P^{X}_{\pm}$,\qquad\begin{quantikz}[thin lines]
& \qw & \qw & \targ{} & \meter{} \\
& \gate{\mathsf{S^{\dagger}}} & \gate{\mathsf{H}} & \ctrl{-1} & \gate{\mathsf{H}} & \gate{\mathsf{S}} & \qw
\end{quantikz}$=P^{Y}_{\pm}$,\qquad\begin{quantikz}[thin lines]
& \qw & \targ{} & \meter{} \\
& \ghost{H} & \ctrl{-1} & \qw
\end{quantikz}$=P^{Z}_{\pm}$
}}

\begin{figure}[tb]
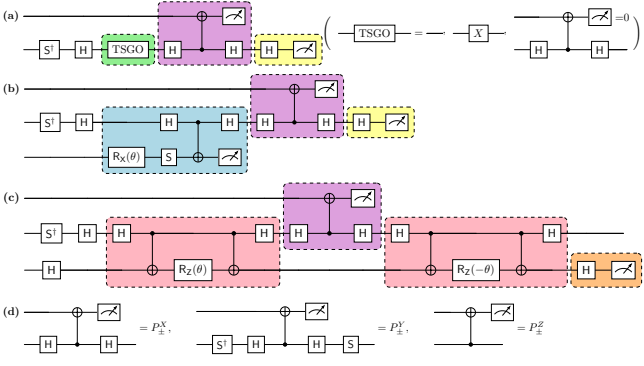

\centering
\begin{tabular}{l}
\usebox\boxD \\
\usebox\boxE \\
\usebox\boxF \\
\usebox\boxG
\end{tabular}
\caption{
Quantum circuits of each method to experimentally evaluate the QRMS disturbance on the quantum computer.
\textbf{(a)} TSM: Three-state generating operations (TSGOs) are implemented by $\{I,X,I+X\}$
\textbf{(b)} WMM: The weak measurement is composed of the $\mathsf{R_{X}}$ gate, $\mathsf{S}$ gate, $\mathsf{H}$ gates, and $\mathsf{CNOT}$ gate.
\textbf{(c)} DEC: The weak interaction $V(\theta)$ is composed of the $\mathsf{R_{Z}}$ gate, $\mathsf{H}$ gates, and $\mathsf{CNOT}$ gates.
\textbf{(d)} The indirect measurement models to realise projective measurements of $X$, $Y$, and $Z$.
}\label{fig:qc}
\end{figure}

\section{\label{sec:results}Results}
\subsection{\label{subsec:theoretical_framework}Theoretical framework}
Let us consider the measurement of an observable $A$ with the arbitrary input state $\rho$ in the measured system $\mathbf{S}$, carried out by the measuring interaction with the probe system $\mathbf{P}$.
The statistical property of this measurement is supposed to be described by a set of measurement operators $\{M_{m}\}$, where the measurement operator $M_{m}$ describes the state-change $\rho \mapsto \rho_{m}$ caused by the measurement and the probability distribution $p(m)$ of the measurement outcome $m$ (see Methods, `Measurement operator formalism').
In this framework, we introduce an additional ancillary system $\mathbf{P'}$ called a weak probe system.
In the following, $X$, $Y$, and $Z$ stand for the Pauli operators, and we consider only finite-dimensional Hilbert spaces for simplicity.

First, we prepare $\mathbf{P'}$ in the eigenstate $|+\rangle$ corresponding to the eigenvalue $+1$ of $X$.
Second, we apply a weak interaction $V(\theta) = \exp(-i \theta B\otimes Z)$, with a coupling strength $\theta$, to the input state $\rho \otimes |+\rangle\langle+|$ of the composite system $\mathbf{S}+\mathbf{P'}$. 
Third, we perform the measurement $\{M_{m}\}$ on $\mathbf{S}$.
Finally, we make the inverse interaction $V^{\dagger}(\theta)$ of $V(\theta)$ act on $\mathbf{S}+\mathbf{P'}$.
In this setting, by approximating up to the quadratic order of $\theta$ $(\ll$$1)$, we calculate the expectation value of the projection observable $P^{X}_{+}:= |+\rangle\langle+|$ in $\mathbf{P'}$ as
\begin{align}
\label{eq:ev_wp}
\langle P^{X}_{+} \rangle^\text{Out}_{\mathbf{P'}} =& \sum_{m} \mathrm{Tr} \left[ ( I \otimes P^{X}_{+} ) V^{\dagger}(\theta) (M_{m}\otimes I) V(\theta) \right.\nonumber\\
&\left.\mbox{}\times ( \rho \otimes |+\rangle\langle+| )V^{\dagger}(\theta) (M_{m}^{\dagger}\otimes I) V(\theta) \right] \nonumber\\
\approx& 1 - \theta^2 \sum_{m} \mathrm{Tr} \left[ [M_{m},B] \rho [M_{m},B]^{\dagger} \right], 
\end{align}
where $\langle \cdot \rangle$ stands for the expectation value in the relevant state.
It should be noted that the expectation value $\langle P^{X}_{+} \rangle^\text{In}_{\mathbf{P'}}$ of $ P^{X}_{+} $ in the input state $|+\rangle$ is equal to $1$ before the disturbance evaluation process.
Moreover, the relative phase in $|+\rangle$, which is the superposition of the computational bases $\{ |0\rangle , |1\rangle \}$, is disturbed by the disturbance evaluation process, leading to the quantitatively detectable decoherence.
This can be expressed as the difference between $\langle P^{X}_{+} \rangle^\text{In}_{\mathbf{P'}}$ and $\langle P^{X}_{+} \rangle^\text{Out}_{\mathbf{P'}}$.
Thus,
\begin{align}
\label{eq:dec_wp}
&\langle P^{X}_{+} \rangle^\text{In}_{\mathbf{P'}}-\langle P^{X}_{+} \rangle^\text{Out}_{\mathbf{P'}} \nonumber\\
&\qquad \approx \theta^2 \sum_{m} \mathrm{Tr} \left[ [M_{m},B] \rho [M_{m},B]^{\dagger} \right]
\end{align}
relates the decoherence in $\mathbf{P'}$, denoted by the left-hand side, to the disturbance with $\theta^{2}$ in $\mathbf{S}$ caused by the measurement, denoted by the right-hand side.
Dividing Eq.~\eqref{eq:dec_wp} by $\theta^2$, we have
\begin{align}
\lim_{\theta \rightarrow 0} \frac{\langle P^{X}_{+} \rangle^\text{In}_{\mathbf{P'}}-\langle P^{X}_{+} \rangle^\text{Out}_{\mathbf{P'}}}{\theta^2} &= \eta^{2}_{\text{O}}(B), \label{eq:dist_dec}
\end{align}
the right-hand side of which is exactly the QRMS disturbance in terms of the density operator \cite{Ozawa05JOB}.
Eq.~\eqref{eq:dist_dec} obviously shows that if the measurement operators $\{M_{m}\}$ commute with the generator $B$ of the weak interactions, the input and non-selective output states coincide with each other.
This is because the application of a pair $V(\theta)$ and its inverse $V^{\dagger}(\theta)$ acts as an identity operation.
However, if $\{M_{m}\}$ and $B$ do not commute, the effect of $V^{\dagger}(\theta)(M_{m}\otimes I)V(\theta)$ presents in the expectation value $\langle P^{X}_{+} \rangle^\text{Out}_{\mathbf{P'}}$ as a function of the commutator $[M_{m},B]$, depending on the magnitude of $\theta$ up to the order of approximation.
This can be interpreted quantitatively as the symmetry breaking between the measurement operator $\{M_{m}\}$ and the generator $B$ of $V(\theta)$.
Thus, the symmetry breaking manifests itself as the decoherence induced in $\mathbf{P'}$, and the magnitude of the decoherence corresponds to the second-order coefficient of the disturbance (see Supplementary Information Sec.~\one~A for the derivation).
The above setup is illustrated in Fig.~\ref{fig:dem}(c) and named here as the \textit{disturbance evaluation circuit} (DEC).

\subsection{\label{subsec:simulation_and_experiment}Simulation and experiment}
\begingroup
\squeezetable
\begin{table*}[tb]
\caption{\label{tab:performance}
Disturbance evaluation performance\footnote[1]{
The projective measurements of $X$, $Y$, and $Z$ are applied to the measured system $\mathbf{S}$, where the input state is $\rho = \ketbra{+i}{+i}$, and the QMRS disturbances $\eta_{\text{O}}(B)$ of $B=X$ are evaluated.
The strength of the weak measurement in the WMM is set to $\theta_{\text{w}} \approx 0.7353 [\text{rad}]$.
The coupling strength of the weak interactions in the DEC are set to $\theta = 0.35 [\text{rad}]$ for the simulation and $\theta = 0.7 [\text{rad}]$ for the experiment.
The evaluation criteria are expressed as Mean $\pm$ SD (Bias, RMSE), where the mean value, standard deviation, bias, and root-mean-square error are considered.
These values are rounded to four significant figures, and the SD, bias, and RMSE are displayed with values aligned to the $10^{-3}$ decimal place due to their small magnitudes, except for the mean.
The mean represents the average value of the QRMS disturbances obtained from $10$ individual simulation and experimental runs, where each run utilises $100,000$ shots.
}
of three methods: the TSM, WMM, and DEC.
}
\begin{ruledtabular}
\begin{tabular}{cccccc}
\multirow{2}{*}{\begin{tabular}{c}Measurement\\Operators\end{tabular}} & \multirow{2}{*}{\begin{tabular}{c}Theoretical\\Values\end{tabular}} & \multirow{2}{*}{\begin{tabular}{c}Evaluation\\Methods\end{tabular}} & \multicolumn{3}{@{}c@{}}{Simulation and Experimental Results: Mean $\pm$ SD$\times10^{-3}$~(Bias$\times10^{-3}$,~RMSE$\times10^{-3}$)} \\
\cmidrule(lr){4-6}%\cmidrule[<wd>](<trim>){a-b}
& & & Noiseless Simulations & Experiments w/o Mitigation & Experiments w/ Mitigation \\
\colrule
\multirow{3}{*}{$P^{X}_{\pm}=(I\pm X)/{2}$} & \multirow{3}{*}{$0$} & TSM & $0.000 \pm 0.000~(0.000,~0.000)$ & $0.1457 \pm 13.62~(145.7,~146.3)$ & $0.08686 \pm 79.35~(86.86,~117.6)$ \\
& & WMM & $\bm{0.06400 \pm 307.9~(64.00,~314.5)}$ & $-0.3760 \pm 266.4~(-376.0,~460.8)$ & $0.02695 \pm 149.5~(26.95,~151.9)$ \\
& & DEC & $0.000 \pm 0.000~(0.000,~0.000)$ & $0.2753 \pm 22.36~(275.3,~276.2)$ & $-0.05254 \pm 64.79~(-52.54,~83.42)$ \\
\multirow{3}{*}{$P^{Y}_{\pm}=(I\pm Y)/{2}$} & \multirow{3}{*}{$2$} & TSM & $1.995 \pm 78.67~(4.600,~78.81)$ & $2.032 \pm 61.75~(32.25,~69.67)$ & $\bm{2.702 \pm 146.3~(702.4,~717.5)}$ \\
& & WMM & $1.995 \pm 161.2~(-5.000,~161.3)$ & $2.017 \pm 358.5~(17.00,~358.9)$ & $1.994 \pm 130.3~(-5.652,~130.4)$ \\
& & DEC & $2.006 \pm 93.79~(5.714,~93.97)$ & $2.014 \pm 54.89~(14.49,~56.77)$ & $2.009 \pm 102.8~(9.002,~103.2)$ \\
\multirow{3}{*}{$P^{Z}_{\pm}=(I\pm Z)/{2}$} & \multirow{3}{*}{$2$} & TSM & $2.003 \pm 41.85~(3.450,~41.99)$ & $2.080 \pm 84.25~(79.90,~116.1)$ & $\bm{2.722 \pm 239.0~(721.6,~760.1)}$ \\
& & WMM & $2.002 \pm 183.0~(2.000,~183.0)$ & $2.077 \pm 352.8~(77.00,~361.2)$ & $2.008 \pm 357.1~(7.501,~357.1)$ \\
& & DEC & $2.007 \pm 51.82~(7.347,~52.33)$ & $2.013 \pm 34.39~(13.06,~36.79)$ & $2.003 \pm 588.0~(2.607,~588.0)$ \\
\end{tabular}
\end{ruledtabular}
\end{table*}
\endgroup

\begingroup
\squeezetable
\begin{table*}[tb]
\caption{\label{tab:comparison}
Comparison of the properties\footnote[2]{We evaluate each item on a three-grade scale: Excellent $\bigcirc$, Good $\triangle$, and Poor $\times$.
The computation cost here means the number of quantum circuit executions required to evaluate the QRMS disturbance. (In other wards, the number of quantum circuits executed on a quantum computer to obtain and calculate classical quantities, i.e., probability distributions, expectation values, etc.)} among three methods: the TSM, WMM, and DEC.
}
\begin{ruledtabular}
\begin{tabular}{ccccccc}
& RMSE (Experiments w/o Mitigation) & \# of gates & \# of measuring devices & Computation cost  & Noise immunity & Mitigation compatibility \\
\colrule
TSM & $\bigcirc$ & $\bigcirc$ & $\bigcirc$ & $\times$ & $\bigcirc$ & $\times$ \\
WMM & $\times$ & $\triangle$ & $\triangle$ & $\bigcirc$ & $\triangle$ & $\bigcirc$ \\
DEC & $\bigcirc$ & $\times$ & $\bigcirc$ & $\bigcirc$ & $\triangle$ & $\bigcirc$
\end{tabular}
\end{ruledtabular}
\end{table*}
\endgroup

We report the results of a simulation and experiment using the \texttt{ibmq\_qasm\_simulator} and \texttt{ibm\_kawasaki}, respectively, via the IBM Quantum Platform \cite{IBMQ} to verify our method and to compare it with the existing methods.
Figs.~\ref{fig:qc}(a), (b), and (c) depict the quantum circuits of the TSM, WMM, and DEC, respectively.
While we have considered an arbitrary measurement of $A$ with measurement operators $\{M_{m}\}$, here we consider the projective measurement of $A$ for which $M_{m}=P^{A}_{m}$.
We also specify $A=X$, $Y$, or $Z$ [see Fig.~\ref{fig:qc}(d)], and $B = X$, while $\rho = \proj{+i}$, where $\ket{+i}$ is the eigenstate of $Y$ corresponing to the eigenvalue $+1$.
We set the coupling strength of the weak interaction $W(\theta_{\text{w}})$ to $\theta_{\text{w}} \approx 0.7353 [\text{rad}]$ for the WMM and the coupling strength of the weak interaction $V(\theta)$ to $\theta = 0.35 [\text{rad}]$ (simulation) and $0.7 [\text{rad}]$ (experiment) for the DEC [see Figs.~\ref{fig:dem}(b) and (c) or Figs.~\ref{fig:qc}(b) and (c)], the total shots to $100,000$, and the number of iterations to $10$.

To compare the performances of the TSM, WMM, and DEC, we consider the mean, standard deviation (SD), bias, and classical root-mean-square error (RMSE) of the obtained QRMS disturbances.
We take the RMSE as the comprehensive evaluation measure (see Methods, `Evaluation criteria' and Supplementary Information Sec.~\two~B for details).

As shown in the Noiseless Simulations and Experiments w/o Mitigation columns of the Table~\ref{tab:performance}, the RMSE of WMM is worse than that of TSM and DEC by an order of magnitude.

Concerning the Noiseless Simulations column, although the theoretical value of the QRMS disturbance for the projective measurement $P^{X}_{\pm}$ of $X$ is $0$, the evaluated value by the WMM deviates from the theoretical one, unlike the TSM and DEC, which yield exactly $0$.
The above deviation from the theoretical value can be attributed to the fact that the TSM and DEC avoid computational errors by exploiting orthogonality and identity, respectively, while the WMM affects the WJD in computational errors due to the perturbation by the weak measurement.

\begin{figure}[tb]%
\centering
\includegraphics[width=.45\textwidth]{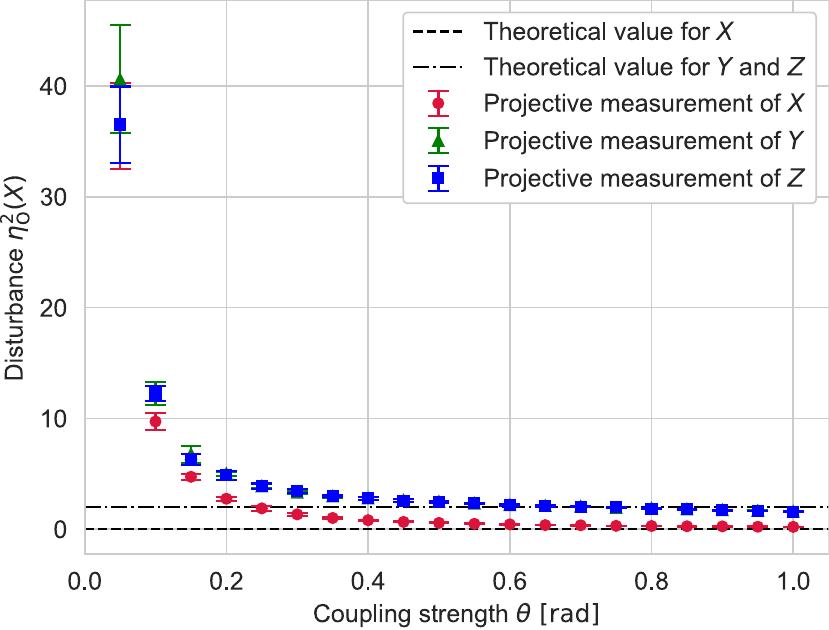}
\caption{\label{fig:dist_theta_extrap_unmiti}
Effect of the coupling strength $\theta$ on the QRMS disturbance $\eta^{2}_{\text{O}}(X)$ evaluated by the DEC in the experiment without mitigation.
The plot shows the variation of the mean of the QRMS disturbances for the projective measurement of $X$, $Y$, and $Z$ when $\theta$ is changed.
The error bars represent the SD.
The evaluation criteria and conditions for obtaining the mean and SD are the same as those described in Table~\ref{tab:performance}.
When using the DEC in the experiment without mitigation, the DEC provides the high and stable performance under the range from $\theta = 0.6[\text{rad}]$ to $\theta = 0.9[\text{rad}]$.
}
\end{figure}

Regarding the Experiments w/o Mitigation column, the results indicate that the noise (e.g., mainly, thermal noise) reduces accuracy in three methods.
In the experiment without mitigation, keeping  $\theta$ set to a small value like $\theta = 0.35 [\text{rad}]$, which is the same with the simulation setting, does not work well for the accurate performance by the DEC, i.e., the evaluated value by the DEC deviates from the theoretical values.
This is because the phase shift due to the weak interactions is sensitive to noise, and dividing the resultant decoherence by $\theta^{2}$ amplifies the effect of the noise.
The reason is supported by the data represented in Fig.~\ref{fig:dist_theta_extrap_unmiti}, where we investigate how the QRMS disturbance obtained by the DEC varies with $\theta$ in noisy experiments.
For a small $\theta$, such as $\theta = 0.05 [\text{rad}]$, the QRMS disturbances and error bars (which represent the SD of each QRMS disturbance) increase; meanwhile, for relatively large $\theta$, the QRMS disturbances approach the theoretical values and the error bars decrease.

\begin{figure}[tb]%
\centering
\includegraphics[width=.45\textwidth]{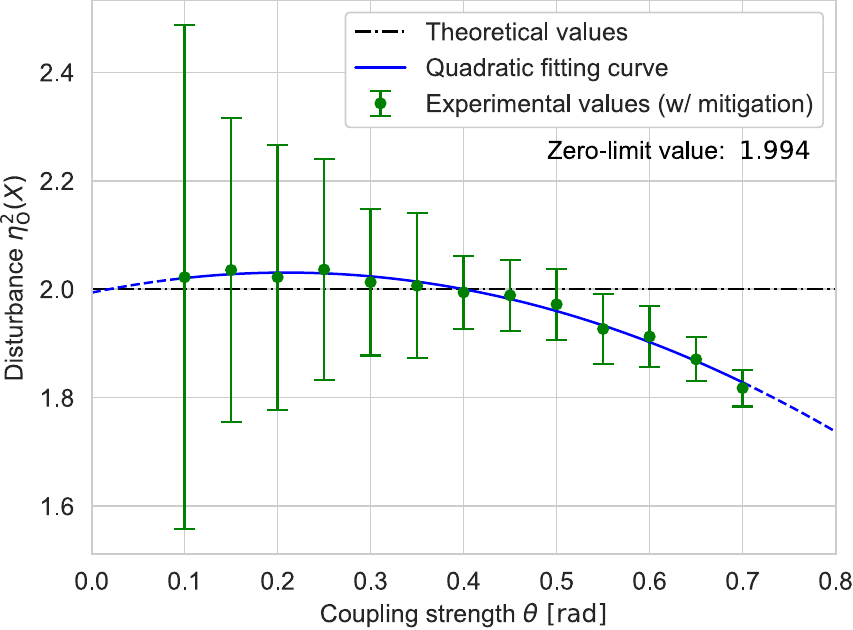}
\caption{\label{fig:dist_theta_extrap_miti}
Effect of the coupling strength $\theta$ on the QRMS disturbance $\eta^{2}_{\text{O}}(X)$ evaluated by the DEC in the experiment with mitigation.
The plot shows the variation of the mean of the QRMS disturbances for the projective measurement of $Z$ when $\theta$ is changed.
The green error bars represent the SD.
The evaluation criteria and conditions for obtaining the mean and SD are the same as those described in Table~\ref{tab:performance}.
The SD decreases with increasing $\theta$, while the evaluated QRMS disturbance tends to deviate from the theoretical value shown by the black dashed line.
It is evident that the intercept (at $\theta \rightarrow 0$) provides the accurate QRMS disturbance by extrapolating and fitting the experimentally obtained results. 
The quadratic function (blue) shows a better fit because the evaluated QRMS disturbance exhibits a cosine-like behaviour as a function of $\theta$.
}
\end{figure}

In the Experiments w/ Mitigation column, both the WMM and DEC more accurately evaluate the QRMS disturbances than that of Experiments w/o Mitigation [see Methods, `Quantum error mitigation (QEM)' and Supplementary Information Sec.~\three~for mitigation techniques].
In particular, the QEM accommodates the WMM and DEC, more than TSM, in the experiment.
While to ensure evaluation stability in noisy experiments, $\theta$ had to be set to $\theta = 0.7 [\text{rad}]$, applying the QEM allows higher evaluation performance closer to the theoretical value even with smaller $\theta$, as low as $\theta = 0.35 [\text{rad}]$, as demonstrated in Fig.~\ref{fig:dist_theta_extrap_miti}.
Furthermore, it is evident that in the limit $\theta\ra0$, the DEC achieves the ideal performance.
This limit is numerically taken by extrapolating the value of the QRMS disturbance with regard to $\theta$ by fitting a quadratic function of $\theta$. 
As for the TSM, the QEM is incapable of effectively reducing the bias in the QRMS disturbance for the projective measurements of $Y$ and $Z$.
To quantify the QRMS disturbance, the TSM needs to take the sum and difference of multiple expectation values, but these computations accumulate the biases of the respective expectation values that could not be fully suppressed.
Consequently, the accuracy of the evaluation deteriorates in total for the TSM with mitigation.

The comprehensive comparison of the properties among the TSM, WMM, and DEC are summarised in Table~\ref{tab:comparison}.
The TSM requires more computation cost than the WMM and DEC, because it is necessary to compute multiple expectation values and to execute several quantum circuit jobs.
In the viewpoint of the accuracy, the TSM and DEC perform better than the WMM
by an order of magnitude.
On the other hand, the WMM has lower computation cost than the TSM and restrains the number of gates compared to the DEC.
The DEC has favorable properties in most aspects except for requiring a larger number of gates than that of the TSM and WMM.
The evaluation method of the QRMS disturbance can be chosen by requirements according to Table~\ref{tab:comparison}.
It is important to note that each of WMM and DEC has the flexibility of choosing the coupling strength, $\theta_{\text{w}}$ or $\theta$.

\section{\label{sec:discussion}Discussion}
%The primary message from the present work is that our method provide the operational viewpoint of the definition of the QRMS disturbance along with the unitarily equivalent relation.
%The fact\md the evaluation of the QRMS disturbance uses the weak interactions, before and after the measurement, acting to the measured system\md implies the unitary equivalence as expressed in the definitions of QRMS disturbance.

We propose the novel method (DEC) and compare it with the existing methods (TSM and WMM) for the experimental evaluation of the QRMS disturbance.
While the WMM and DEC are similar in the sense that they both employ the weak interactions, they differ in how they store the information of the observable correlation in the WJD used in the theory of quantum perfect correlation \cite{Ozawa05PLA,Ozawa06} and how they utilise the weak probe system.
The WMM stores the information of the observable correlation in a \ti{classical register} through the weak measurement to obtain the QRMS disturbance via the WJD.
On the other hand, the DEC stores the observable correlation in a \ti{quantum register} through the weak interactions to obtain the QRMS disturbance via the second-order coefficient of the decoherence.
(The qubit state of the weak probe system can be viewed as a quantum memory to store the quantum information.)
A detailed explanation of how to interpret the decoherence can be found in Supplementary Information Sec.~\one~B, where we utilise a $\ell_{1}$ norm coherence measure \cite{Baumgratz14}.

\section{\label{sec:methods}Methods}
\subsection{\label{subsec:error_disturbance}Quantum root-mean-square error and disturbance}
\subsubsection{\label{subsubsec:noise_disturbance_operators_formalism}Noise and disturbance operators formalism}
Consider a measurement of an observable $A$ in the measured system $\mathbf{S}$, described by a Hilbert space $\mathcal{H}$ with an input state $\rho$, carried out by the measuring interaction with the probe system $\mathbf{P}$.
%The statistical properties of the measurement are fully characterised by the quantum instrument $\mathcal{I}$ \cite{Davies70,Ozawa84}.
%The quantum instrument is specifically modelled as a measuring process;
The measuring process of this measurement is specified by
%The resulting model is called an indirect measurement model.
%The indirect measurement model $\mathbf{M}$ is represented by 
a quadruple $(\mathcal{K},\sigma,U,M)$, called an indirect measurement model, consisting of a Hilbert space $\mathcal{K}$, a density operator $\sigma$ on $\mathcal{K}$, a unitary operator $U$ on $\mathcal{H}\otimes\mathcal{K}$, and an observable (self-adjoint operator) $M$ on $\mathcal{K}$, where the Hilbert space $\mathcal{K}$ describes the state space of a probe system $\mathbf{P}$,
the density operator $\sigma$ describes the initial state of $\mathbf{P}$, the unitary operator $U$ describes the time evolution of the composite system $\mathbf{S}+\mathbf{P}$ during the measuring interaction from $t=0$ to $t=\tau$, and the observable $M$ describes the meter observable to be read out at the instant $t=\tau$ just after the measuring interaction.
In the Heisenberg picture, we introduce the Heisenberg operators $A(0)=A \otimes I$, $A(\tau)=U^{\dagger}(A \otimes I)U$, $M(0)=I\otimes M$, and $M(\tau)=U^{\dagger}(I \otimes M)U$, where $I$ is an identity operator.
The outcome of this measurement is obtained by the measurement of $M(\tau)$ in the state $\rho\otimes\sigma$.
For an arbitrary observable (disturbed observable) $B$ of $\mathbf{S}$, the measuring interaction changes $B(0)=B \otimes I$ to $B(\tau)=U^{\dagger}(B \otimes I)U$.
Then, the (operator-based) QRMS error $\varepsilon_{\text{O}}(A)$ and the (operator-based) QRMS disturbance $\eta_{\text{O}}(B)$ in the state $\rho$ are defined as follows:
\begin{align}
\varepsilon^{2}_{\text{O}}(A) &:= \langle E^{2}(A) \rangle = \left\| [M(\tau) - A(0)] (\sqrt{\rho}\otimes\sqrt{\sigma}) \right\|^{2}_{\text{HS}}, \label{eq:error}\\
\eta^{2}_{\text{O}}(B) &:= \langle D^{2}(B) \rangle = \left\| [B(\tau) - B(0)] (\sqrt{\rho}\otimes\sqrt{\sigma}) \right\|^{2}_{\text{HS}}, \label{eq:dist}
\end{align}
where $E(A)=M(\tau) - A(0)$ is the error operator, $D(B)=B(\tau) - B(0)$ is the disturbance operator, and $\|\cdot\|_{\text{HS}}$ denotes the Hilbert\nd Schmidt norm  \cite{Ozawa03,Ozawa04}.

\subsubsection{\label{subsubsec:measurement_operator_formalism}Measurement operator formalism}
Assume that the quantum instrument $\{\mathcal{I}_{m}\}$ is given by $\mathcal{I}_{m} (\cdot) = M_{m} (\cdot) M^{\dagger}_{m}$, where $\{M_{m}\}$ is a family of measurement operators \cite{Nielsen10}.
The probability of obtaining an measurement outcome $m$ is given by $p(m) = \Tr\left[M^{\dagger}_{m}M_{m}\rho\right]$, and the input state $\rho$ changes to the output state $\rho_{m}=M_{m}\rho M^{\dagger}_{m}/\Tr\left[M^{\dagger}_{m}M_{m}\rho\right]$ for the measurement outocome $m$ after the measurement.
The operators defined by $\Pi_{m}:= M^{\dagger}_{m}M_{m}$ satisfy the conditions $\Pi_{m} \ge 0 $ for all $m$ and $\sum_{m}\Pi_{m} = I$; the family $\{\Pi_{m}\}$ is called a probability operator-valued measure (POVM), and it determines the probability distribution of the measurement outcome for the input state. 
Using the measurement operators, Eqs.~\eqref{eq:error} and \eqref{eq:dist} can be rewritten as
\begin{align}
\varepsilon^{2}_{\text{O}}(A) &= \sum_{m}\left\| M_{m}(m-A) \sqrt{\rho} \right\|^2_{\text{HS}}, \label{eq:error_using_mops}\\
\eta^{2}_{\text{O}}(B) &= \sum_{m}\left\| [M_{m},B] \sqrt{\rho} \right\|^2_{\text{HS}}. \label{eq:dist_using_mops}
\end{align}
The $n$-th moment operator of $\{\Pi_{m}\}$ is defined by $O_{A}^{(n)}:=\sum_{m}m^{n}\Pi_{m}$.
If $\{\Pi_{m}\}$ is a projection-valued measure (PVM), we can simplify Eq.~\eqref{eq:error_using_mops} to $\varepsilon_{\text{O}}(A) = \| (O_{A}-A) |\psi\rangle \|_{\text{HS}}$ by setting $O_{A}=O_{A}^{(1)}$ \cite{Ozawa05JOB}.

\subsection{\label{subsec:previous_evaluation_methods}Existing evaluation methods for error and disturbance}
\subsubsection{\label{subsubsec:tsm}Three-state method (TSM)}
We can obtain an equivalent expression of the QRMS disturbance using $O_{B} := \sum_{m} M^{\dagger}_{m}BM_{m}$ and $O^{(2)}_{B} := \sum_{m} M^{\dagger}_{m}B^{2}M_{m}$ as follows:
\begin{align}
\label{eq:dist_tsm}
\eta_{\text{O}}^{2}(B) =& \langle\psi|B^{2}|\psi\rangle + \langle\psi|O^{(2)}_{B}|\psi\rangle + \langle\psi|O_{B}|\psi\rangle \nonumber\\
&+ \langle B\psi|O_{B}|B\psi\rangle - \langle(B+I)\psi|O_{B}|(B+I)\psi\rangle.
\end{align}
This expression implies that the QRMS disturbance can be obtained by measuring the first and second moments of the output from the second apparatus measuring $B$, whose input is the output of the first apparatus described by $\{M_{m}\}$, for three different input states $|\psi_{1}\rangle := |\psi\rangle$, $|\psi_{2}\rangle := B |\psi\rangle/\|B |\psi\rangle\|$, and $|\psi_{3}\rangle := (B+I)|\psi\rangle/\|(B+I)|\psi\rangle\|$.
Similarly, the same expression can be obtained for the QRMS error by replacing $B$ and $O_{B}$ in Eq.~\eqref{eq:dist_tsm} with $A$ and $O_{A}$, respectively \cite{Ozawa04}.

As an example, the QRMS disturbance of $B=X$ in the input state $\ket{\psi}$ is of the form
\begin{align}
\label{eq:dist_x_tsm}
\eta_{\text{O}}^{2}(X) =& 2 + \langle\psi|O_{X}|\psi\rangle + \langle \psi|XO_{X}X|\psi\rangle \nonumber\\
&- 4 \langle\psi|(|+\rangle\langle+|)O_{B}(|+\rangle\langle+|)|\psi\rangle. 
\end{align}

\subsubsection{\label{subsubsec:wmm}Weak measurement method (WMM)}
We can also write the QRMS disturbance using the family $\{P^{B}_{b}\}$ of spectral projections of $B=\sum_{b}bP^{B}_{b}$ as follows:
\begin{align}
\eta_{\text{O}}^{2}(B)&=\left\| [ B(\tau) - B(0) ] (\sqrt{\rho}\otimes\sqrt{\sigma}) \right\|^{2}_{\text{HS}}\nonumber\\
&=\sum_{b_{0},b_{\tau}}(b_{\tau}-b_{0})^{2} \mathrm{Re} \langle U^{\dagger} (P^{B}_{b_{\tau}}\otimes I) U (P^{B}_{b_{0}}\otimes I) \rangle\nonumber\\
&=\sum_{b_{0},b_{\tau}}(b_{\tau}-b_{0})^{2}\mathrm{Re}p_{\text{w}}(b_{0},b_{\tau}), \label{eq:dist_wmm}
\end{align}
where $U$ is the measuring interaction and $\sigma$ is the initial state of $\mathbf{P}$ in the indirect measurement model $(\mathcal{K},\sigma,U,M)$.
$p_{\text{w}}(b_{0},b_{\tau})$ is the weak joint distribution (WJD) \cite[Ch.~5]{Ozawa05PLA} defined by
\begin{align}
p_{\text{w}}(b_{i},b_{f})&:= \mathrm{Tr}\left[ ( P^{B}_{b_{f}}\otimes I) U (P^{B}_{b_{i}}\otimes I) (\rho \otimes \sigma) U^{\dagger} \right]\nonumber\\
&=\langle U^{\dagger} (P^{B}_{b_{f}}\otimes I) U (P^{B}_{b_{i}}\otimes I) \rangle.
\end{align}
The weak value $\langle P^{B}_{b_{i}}\rangle_{\text{w}}$ \cite{Aharonov88} is defined by
\begin{equation}
\langle P^{B}_{b_{i}}\rangle_{\text{w}} := \frac{\mathrm{Tr}\left[(P^{B}_{b_{f}}\otimes I) U (P^{B}_{b_{i}}\otimes I) (\rho \otimes \sigma) U^{\dagger} \right]}{\mathrm{Tr}\left[ (P^{B}_{b_{f}}\otimes I) U (\rho \otimes \sigma) U^{\dagger} \right]}. \label{eq:wv}
\end{equation}
This can be obtained by the state preparation (pre-selection) of $\rho$, the weak measurement of $P^{B}_{b_{i}}$, and the strong measurement (post-selection) of $P^{B}_{b_{f}}$.
From Eq.~\eqref{eq:wv}, the weak value $\langle P^{B}_{b_{i}}\rangle_{\text{w}}$ of the projection $P^{B}_{b_{i}}$ is interpreted as the weak conditional probability $p_{\text{w}}(b_{i}|b_{f})$ of the event $P^{B}_{b_{i}}$, conditional upon the event $P^{B}_{b_{f}}$ in the input state $\rho$ \cite{Steinberg95PRA,Steinberg95PRL}, where $p_{\text{w}}(b_{i}|b_{f})=p_{\text{w}}(b_{i},b_{f})/p(b_{f})$. 
Therefore, the WJD $p_{\text{w}}(b_{i},b_{f})$ can be expressed as
\begin{align}
p_{\text{w}}(b_{i},b_{f})=\langle P^{B}_{b_{i}}\rangle_{\text{w}}p(b_{f})
\end{align}
and experimentally obtained by the weak measurement, so that the QRMS disturbance can be evaluated by the weak measurement \cite{Lund10}.
The QRMS error for $A$ can also be obtained by a procedure similar to the above for the QRMS disturbance \cite{Ozawa05PLA,Lund10}.

In the case $B=X$, the weak measurement of $P^{X}_{x_{i}}$ (corresponding to $P^{B}_{b_{i}}$) used in the WMM is represented by the POVM $\{\Pi^{X}_{\pm}\}$ such that
\begin{align}
\Pi^{X}_{\pm}=\frac{1}{2}(I\pm \cos2\theta_{\text{w}}X),
\end{align}
which is implemented by the weak interaction $W(\theta_{\text{w}})$ given by
\begin{align}
W(\theta_{\text{w}})&=\mathsf{H}_{\mathbf{S}}\circ\mathsf{CNOT}_{\mathbf{SP'}}\circ\mathsf{H}_{\mathbf{S}}\circ\mathsf{S}_{\mathbf{P'}}\circ\mathsf{R_{X}}(2\theta_{\text{w}})_{\mathbf{P'}}\nonumber\\
&=\proj{+}\otimes e^{-i\frac{\pi}{4}Z}e^{-i\theta_{\text{w}}X}\nonumber\\
&\qquad\qquad\mbox{}+\proj{-}\otimes Xe^{-i\frac{\pi}{4}Z}e^{-i\theta_{\text{w}}X}, \label{eq:W}
\end{align}
where the coupling strength $\theta_{\text{w}}$ is assumed to be small [see Figs.~\ref{fig:dem}(b) and (e)].
In this case, the QRMS disturbance of $X$ is expressed in a concise form \cite{Lund10,Kaneda14}
\begin{equation}
\eta^{2}_{\text{O}}(X)=2\left( 1 - \lim_{\theta_{\text{w}}\ra \frac{\pi}{4}}\sum_{x_{i},x_{f}=\pm1}\frac{1}{\cos 2\theta_{\text{w}}} x_{i} x_{f} p(x_{i}, x_{f})\right). \label{eq:dist_x_wmm}
\end{equation}
Here
\begin{align}
p(x_{i},x_{f})&=\sum_{m}p(m,x_{i},x_{f}), \\
\lefteqn{p(m,x_{i},x_{f})}\hspace{3.6em}\nonumber\\
&\hspace{-2em}=\Tr\left[(P^{M}_{m}\otimes P^{X}_{f}\otimes P^{X}_{i})(U\otimes I_{\mathbf{P'}})(I_{\mathbf{P}}\otimes W(\theta_{\text{w}}))\right.\nonumber\\
&\hspace{-2em}\left.\quad\mbox{}\times(\sigma\otimes\rho\otimes\proj{0})(I_{\mathbf{P}}\otimes W^{\dagger}(\theta_{\text{w}}))(U^{\dagger}\otimes I_{\mathbf{P'}})\right].
\end{align}
The coupling strength is set to $\theta_{\text{w}} \approx 0.7353 [\text{rad}]$ such that the measurement strength is $\cos2\theta_{\text{w}} \approx 0.1$.

\subsection{\label{subsec:simulation_details}Simulation and experiment details}
In this study, we utilise Qiskit as the frontend and employ the \texttt{ibmq\_qasm\_simulator} and \texttt{ibm\_kawasaki} as the backends \cite{IBMQ}.
Additionally, we assign the quantum register as $[\text{Q}66,\text{Q}73,\text{Q}85]$.

\subsubsection{\label{subsubsec:evaluation_criteria}Evaluation criteria}
The criteria established to investigate the performance of the disturbance evaluation methods are defined as follows:
\ti{Total shots}: the total number of measurements (sample size) for each measuring device;
\ti{Iterations}: the number of trials (simulation and/or experiment runs) required to obtain disturbances from the measurement results of total shots;
\ti{Mean} $m = \frac{1}{n} \sum_{j=1}^{n}a_j$: the average value of disturbances $\{a_j\}$ obtained from $n$ iterations;
\ti{Standard deviation} (SD) $\sigma = \left[\frac{1}{n} \sum_{j=1}^{n} (a_j-m)^2 \right]^{1/2}$: the degree of variability in the $n$ obtained disturbances;
\ti{Bias} $b = m-a$: the difference between the mean and the theoretical value $a$;
\ti{Root-mean-square error} (RMSE) $e = \left[\frac{1}{n} \sum_{j=1}^{n}(a_j-a)^2 \right]^{1/2}$: the sum $e^{2} = \sigma^{2} + b^{2}$ of the squares of the SD and bias, representing the overall performance of the evaluation method.
More information is given in Supplementary Information Sec.~\two.

\subsubsection{\label{subsubsec:qem}Quantum error mitigation (QEM)}
Current quantum computers are noisy intermediate-scale quantum (NISQ) devices, which lack error-correction capabilities.
Therefore, it is necessary to effectively utilise the QEM to \textit{mitigate} noise (sources of error) that occurs during the computation process \cite{Endo21}.
The QEM effectively reduces errors through the following process: highlighting the statistical properties of errors, modifying the parts of the original quantum algorithm (at the software level) or gate operation (at the hardware level) that need to be compensated for, and applying existing classical information processing to the measured results as post-processing.
This allows us to obtain values that are close to the noiseless ideal expectation values \cite{Kandala19,Kim23}.

We implement relatively simple and effective QEM techniques known as readout error mitigation (REM) \cite{Chen19,Sarovar20,Bravyi21} for measurement errors and zero-noise extrapolation (ZNE) \cite{Li17,Temme17,Dumitrescu18} for gate errors in the experiments.
REM is a method that performs quantum detector tomography on the measuring devices to obtain a confusion matrix (calibration matrix) and calculates its inverse using least-squares or pseudoinverse methods to compensate for bit-flip probabilities.
ZNE estimates error-free computational results by extrapolating the noise-scaled results after applying a noise-scaling process that gradually increases errors in the raw output of quantum computation.
The details of the QEM implementations in the experiments are provided in Supplementary Information Sec.~\three.

\section{\label{sec:data_availability}Data availability}
The data generated in this study are available from the corresponding  author upon reasonable request.

\section{\label{sec:code_availability}Code availability}
The code used in this study is available from the corresponding author upon reasonable request.

\bibliography{ref.bib}

\section{\label{sec:acknowledgments}Acknowledgments}
We thank Yasuhito Kawano for useful discussions.
H.E. is grateful to Masahide Sasaki for an advice on this research direction.
This work was supported by JSPS KAKENHI Grant Numbers JP24H01566, JP22K03424, JP21K11764, JP21K04915, and by JST CREST Grant Number JPMJCR23P4, Japan.
This work was partly supported by UTokyo Quantum Initiative.
H.E. was partly supported by a research assistantship at the Quantum ICT Collaboration Center, NICT.

\section{\label{sec:author_contributions}Author contributions}
H.E. conducted the theoretical analysis with advice from M.O., and performed the experiment with advice from A.T., while M.O. and A.T. provided scientific guidance on theoretical and experimental questions and tasks.
All authors contributed to the writing of the manuscript.

\section{\label{sec:competing_interests}Competing interests}
The authors declare no competing interests.

\section{\label{sec:additional_information}Additional information}
\noindent
\tb{Extended data} are not included in this paper.

\noindent
\tb{Supplementary information} is included in this study.

\noindent
\tb{Correspondence and requests for materials} should be addressed to Haruki Emori.

%% file: supp.tex
\begin{center}
{\large \bf Supplementary Information for \protect\\ 
`Disturbance Evaluation Circuit in Quantum Measurement'}\\
\vspace*{0.3cm}
Haruki Emori$^{1,3}$, Masanao Ozawa$^{2,3,4}$, and Akihisa Tomita$^{1}$\\
\vspace*{0.1cm}
$^{1}${\small \em Graduate School of Information Science and Technology,\\ Hokkaido University, Kita-ku, Sapporo, Hokkaido 060-0814, Japan}\\
$^{2}${\small \em Center for Mathematical Science and Artificial Intelligence,\\ Chubu University Academy of Emerging Sciences,\\Chubu University, Matsumoto-cho, Kasugai, Aichi 487-8501, Japan}\\
$^{3}${\small \em RIKEN Innovation Design Office, Hirosawa, Wako, Saitama 351–0198, Japan}\\
$^{4}${\small \em Graduate School of Informatics, Nagoya University, Chikusa-ku, Nagoya, Aichi 464-8601, Japan}\\
(Dated: \today)
\end{center}

\setcounter{equation}{0}
\setcounter{lemma}{1}
\setcounter{page}{1}
\renewcommand{\theequation}{S.\arabic{equation}}

\vspace{-5.5em}

\begin{myquote}{0.55in}
\quad The Supplementary Information is organised as follows.
In Sec.~\one, we provide the derivation details of the DEC and elucidate the mechanism of it.
In Sec.~\two, we specifically explain the conditions and evaluation criteria that we set for the simulation and experiment.
In Sec.~\three, we introduce the theory and implementation of the QEMs (REM and ZNE) used to suppress noise for the noisy experiment.
\end{myquote}

%\vspace{+5.5em}

\section{\label{s.sec:ta_dec}\one. Theoretical analysis of disturbance evaluation circuit}
\subsection{A. Derivation details}
Here, we describe in detail the dynamics of the setup shown in Fig.~\ref{fig:dem}(c).
The sequential operations of the weak interaction $V(\theta)$, the measurement $\{M_{m}\}$, and the inverse weak interaction $V^{\dagger}(\theta)$ can be written as
\begin{equation}
\begin{split}
\label{eq:weak_unitary_meas_op}
\lefteqn{V^{\dagger}(\theta) (M_{m}\otimes I) V(\theta)}\quad\quad\\
&= \exp \left( i \theta B \otimes Z \right) (M_{m}\otimes I) \exp \left(-i \theta B \otimes Z \right) \\
&= \Bigl( I\otimes I + i \theta B \otimes Z - \frac{1}{2} \theta^{2} B^{2} \otimes I + O(\theta^{3}) \Bigr) (M_{m}\otimes I) \Bigl( I\otimes I -i \theta B \otimes Z - \frac{1}{2} \theta^{2} B^{2} \otimes I + O(\theta^{3}) \Bigr) \\
&= \Bigl( M_{m}\otimes I + i \theta B M_{m} \otimes Z - \frac{1}{2} \theta^{2} B^{2} M_{m} \otimes I + O(\theta^{3}) \Bigr) \Bigl( I\otimes I -i \theta B \otimes Z - \frac{1}{2} \theta^{2} B^{2} \otimes I + O(\theta^{3}) \Bigr) \\
&= M_{m}\otimes I + i \theta B M_{m} \otimes Z - \frac{1}{2} \theta^{2} B^{2} M_{m} \otimes I - i \theta M_{m} B \otimes Z + \theta^{2} B M_{m} B \otimes I - \frac{1}{2} \theta^{2} M_{m} B^{2} \otimes I + O(\theta^{3}) \\
&= M_{m}\otimes I - i \theta [M_{m},B] \otimes Z + \frac{\theta^{2}}{2}[B,[M_{m},B]] \otimes I + O(\theta^{3}),
\end{split}
\end{equation}
where $\otimes$ stands for the tensor product of operators for $\mathbf{S}$ and $\mathbf{P'}$ in this order, e.g., $I\otimes I=I_{\mathbf{S}}\otimes I_{\mathbf{P'}}$.
Similary, we have
\begin{equation}
\label{eq:weak_unitary_meas_op_dagger}
V^{\dagger}(\theta) (M^{\dagger}_{m}\otimes I) V(\theta) = M^{\dagger}_{m}\otimes I - i \theta [M^{\dagger}_{m},B] \otimes Z + \frac{\theta^{2}}{2}[B,[M^{\dagger}_{m},B]] \otimes I + O(\theta^{3}).
\end{equation}
With the help of Eqs.~\eqref{eq:weak_unitary_meas_op} and \eqref{eq:weak_unitary_meas_op_dagger}, we can calculate the expectation value of $P^{X}_{+}$ in the output state up to the second order of $\theta$:
\begin{equation}
\begin{split}
\langle P^{X}_{+} \rangle^\text{Out}_{\mathbf{P'}} &:= \sum_{m} \mathrm{Tr} \Bigl[ ( I \otimes  P^{X}_{+}  ) V^{\dagger}(\theta) (M_{m}\otimes I) V(\theta) ( \rho \otimes  \proj{+}  ) V^{\dagger}(\theta) (M_{m}^{\dagger}\otimes I) V(\theta) \Bigr] \\
&= \sum_{m} \mathrm{Tr} \Bigl[ ( I \otimes  P^{X}_{+}  ) (M_{m}\otimes I - i \theta [M_{m},B] \otimes Z + \frac{\theta^{2}}{2}[B,[M_{m},B]] \otimes I + O(\theta^{3})) \Bigr.\\
&\Bigl. \quad \mbox{}\times ( \rho \otimes \proj{+} ) (M^{\dagger}_{m}\otimes I - i \theta [M^{\dagger}_{m},B] \otimes Z + \frac{\theta^{2}}{2}[B,[M^{\dagger}_{m},B]] \otimes I + O(\theta^{3})) \Bigr] \\
&= \sum_{m} \mathrm{Tr} \Bigl[ ( I \otimes  P^{X}_{+}  ) ( M_{m} \rho M^{\dagger}_{m} \otimes |+\rangle\langle+| - i \theta [M_{m},B] \rho M^{\dagger}_{m} \otimes |-\rangle\langle+| \Bigr.\\
&\Bigl. \quad \mbox{}- i \theta M_{m} \rho [M^{\dagger}_{m},B] \otimes |+\rangle\langle-| - \theta^2 [M_{m},B] \rho [M^{\dagger}_{m},B] \otimes |-\rangle\langle-| \Bigr.\\
&\Bigl. \quad \mbox{}+ \frac{\theta^{2}}{2} [B,[M_{m},B]] \rho M^{\dagger}_{m} \otimes |+\rangle\langle+| + \frac{\theta^{2}}{2} M_{m} \rho [B,[M^{\dagger}_{m},B]] \otimes |+\rangle\langle+| + O(\theta^{3})) \Bigr] \\
&= \sum_{m} \mathrm{Tr} \Bigl[ M_{m} \rho M^{\dagger}_{m} + \frac{\theta^{2}}{2} [B,[M_{m},B]] \rho M^{\dagger}_{m} + \frac{\theta^{2}}{2} M_{m} \rho [B,[M^{\dagger}_{m},B]] \Bigr] + O(\theta^{3})\\
&= 1 + \theta^2 \sum_{m} \mathrm{Tr} \left[ [M_{m},B] \rho [M^{\dagger}_{m},B] \right] + O(\theta^{3})\\
&= 1 - \theta^2 \sum_{m} \mathrm{Tr} \left[ [M_{m},B] \rho [M_{m},B]^{\dagger} \right] + O(\theta^{3}).
\end{split}
\end{equation}
Using this result, we can express the decoherence in the weak probe as follows:
\begin{equation}
\begin{split}
\label{eq:decoherence}
\langle P^{X}_{+} \rangle^\text{In}_{\mathbf{P'}}-\langle P^{X}_{+} \rangle^\text{Out}_{\mathbf{P'}} &= 1-\langle P^{X}_{+} \rangle^\text{Out}_{\mathbf{P'}} \\
&= \theta^2 \sum_{m} \mathrm{Tr} \left[ [M_{m},B] \rho [M_{m},B]^{\dagger} \right] + O(\theta^{3}).
\end{split}
\end{equation}
Since the right-hand side of Eq.~\eqref{eq:decoherence} is $\theta^{2}$ times the QRMS disturbance \cite{Ozawa03,Ozawa04},
the QRMS disturbance can be obtained by dividing the right-hand side of Eq.~\eqref{eq:decoherence} by $\theta^{2}$ and taking the limit as $\theta\ra0$:
\begin{equation}
\begin{split}
\label{s.eq:dist_p}
\lim_{\theta \rightarrow 0} \frac{\langle P^{X}_{+} \rangle^\text{In}_{\mathbf{P'}}-\langle P^{X}_{+} \rangle^\text{Out}_{\mathbf{P'}}}{\theta^{2}} &= \sum_{m} \mathrm{Tr} \left[ [M_{m},B] \rho [M_{m},B]^{\dagger} \right] \\
&=\eta^{2}_{\text{O}}(B).
\end{split}
\end{equation}

In the DEC introduced above, the QRMS disturbance is obtained as the second-order derivative of the decoherence induced in a newly
introduced weak probe system $\mathbf{P'}$, with respect to the
coupling strength $\theta$ of the weak interaction $V(\theta)$ at $\theta=0$.
This second-order coefficient of the decoherence is calculated from the difference between the expectation values of $ P^{X}_{+} $ in the input and output states.
Alternatively, the same result can be obtained by calculating the difference between the expectation values of $X$ in the input and output states, as in the setup in~\cite{Hofmann21}.
If we adopt it, the factor on the QRMS disturbance is doubled:
\begin{equation}
\begin{split}
\langle X \rangle^{\text{Out}}_{\mathbf{P'}} &= \sum_{m} \mathrm{Tr} \Bigl[ ( I \otimes X ) V^{\dagger}(\theta) (M_{m}\otimes I) V(\theta) ( \rho \otimes  \proj{+} ) V^{\dagger}(\theta) (M_{m}^{\dagger}\otimes I) V(\theta) \Bigr] \\
&= \sum_{m} \mathrm{Tr} \Bigl[ M_{m} \rho M^{\dagger}_{m} + \frac{\theta^{2}}{2} [B,[M_{m},B]] \rho M^{\dagger}_{m} + \frac{\theta^{2}}{2} M_{m} \rho [B,[M^{\dagger}_{m},B]] \Bigr.\\
&\Bigl.\quad \mbox{}+ \theta^2 [M_{m},B] \rho [M^{\dagger}_{m},B] \Bigr] + O(\theta^{3}) \\
&= 1 + 2 \theta^2 \sum_{m} \mathrm{Tr} \left[ [M_{m},B] \rho [M^{\dagger}_{m},B] \right] + O(\theta^{3}) \\
&= 1 - 2 \theta^2 \sum_{m} \mathrm{Tr} \left[ [M_{m},B] \rho [M_{m},B]^{\dagger} \right] + O(\theta^{3}).
\end{split}
\end{equation}
Therefore, it is necessary to divide $\langle X \rangle^{\text{In}}_{\mathbf{P'}}-\langle X \rangle^{\text{Out}}_{\mathbf{P'}}$ by $2 \theta^2$ for the evaluation of the QRMS disturbance:
\begin{equation}
\begin{split}
\label{s.eq:dist_x}
\eta^{2}_{\text{O}}(B) = \lim_{\theta \rightarrow 0} \frac{\langle X \rangle^{\text{In}}_{\mathbf{P'}}-\langle X \rangle^{\text{Out}}_{\mathbf{P'}}}{2 \theta^{2}}.
\end{split}
\end{equation}
The above calculations use the following facts:
\begin{equation}
\sum_{m} \mathrm{Tr} \Bigl[ \frac{\theta^{2}}{2} [B,[M_{m},B]] \rho M^{\dagger}_{m} + \frac{\theta^{2}}{2} M_{m} \rho [B,[M^{\dagger}_{m},B]] \Bigr] = \theta^2 \sum_{m} \mathrm{Tr} \left[ [M_{m},B] \rho [M^{\dagger}_{m},B] \right],
\end{equation}
where
\begin{equation}
\begin{split}
&\sum_{m} \mathrm{Tr} \Bigl[ \frac{\theta^{2}}{2} [B,[M_{m},B]] \rho M^{\dagger}_{m} + \frac{\theta^{2}}{2} M_{m} \rho [B,[M^{\dagger}_{m},B]] \Bigr] \\
&= \frac{\theta^{2}}{2} \sum_{m} \mathrm{Tr} \bigl[ [ B ( M_{m} B - B M_{m} ) - ( M_{m} B - B M_{m} ) B ] \rho M^{\dagger}_{m} \bigr.\\
&\bigl. \quad \mbox{}+  M_{m} \rho [ B ( M^{\dagger}_{m} B - B M^{\dagger}_{m} ) - ( M^{\dagger}_{m} B - B M^{\dagger}_{m} ) B ] \bigr] \\
&= \frac{\theta^{2}}{2} \sum_{m} \mathrm{Tr} \bigl[ B M_{m} B \rho M^{\dagger}_{m} - B^{2} M_{m} \rho M^{\dagger}_{m} - M_{m} B^{2} \rho M^{\dagger}_{m} + B M_{m} B \rho M^{\dagger}_{m} \bigr.\\
&\bigl. \quad \mbox{}+ M_{m} \rho B M^{\dagger}_{m} B - M_{m} \rho B^{2} M^{\dagger}_{m} - M_{m} \rho M^{\dagger}_{m} B^{2} + M_{m} \rho B M^{\dagger}_{m} B \bigr] \\
&= \frac{\theta^{2}}{2} \sum_{m} \mathrm{Tr} \bigl[ 2 B M_{m} B \rho M^{\dagger}_{m} - B^{2} M_{m} \rho M^{\dagger}_{m} - M_{m} B^{2} \rho M^{\dagger}_{m} \bigr.\\
&\bigl. \quad \mbox{}+ 2 M_{m} \rho B M_{m}^{\dagger} B - M_{m} \rho B^{2} M_{m}^{\dagger} - M_{m} \rho M_{m}^{\dagger} B^{2} \bigr] \\
&= \theta^{2} \sum_{m} \Bigl( \mathrm{Tr} \left[ B M_{m} B \rho M_{m}^{\dagger} \right] + \mathrm{Tr} \left[ B M_{m} \rho B M_{m}^{\dagger} \right] - \mathrm{Tr} \left[ B^{2} M_{m} \rho M_{m}^{\dagger} \right] \Bigr) - \theta^{2} \mathrm{Tr} \left[ B^{2} \rho \right]
\end{split}
\end{equation}
and
\begin{equation}
\begin{split}
&\theta^2 \sum_{m} \mathrm{Tr} \left[ [M_{m},B] \rho [M^{\dagger}_{m},B] \right] \\
&= \theta^2 \sum_{m} \mathrm{Tr} \left[ ( M_{m} B - B M_{m} ) \rho ( M^{\dagger}_{m} B -B M^{\dagger}_{m} )\right] \\
&= \theta^2 \sum_{m} \mathrm{Tr} \left[ ( M_{m} B \rho - B M_{m} \rho )  ( M^{\dagger}_{m} B -B M^{\dagger}_{m} )\right] \\
&= \theta^2 \sum_{m} \mathrm{Tr} \left[ M_{m} B \rho  M^{\dagger}_{m} B - B M_{m} \rho M^{\dagger}_{m} B - M_{m} B \rho B M^{\dagger}_{m} + B M_{m} \rho B M^{\dagger}_{m} \right] \\
&= \theta^{2} \sum_{m} \Bigl( \mathrm{Tr} \left[ B M_{m} B \rho M^{\dagger}_{m} \right] + \mathrm{Tr} \left[ B M_{m} \rho B M^{\dagger}_{m} \right] - \mathrm{Tr} \left[ B^{2} M_{m} \rho M^{\dagger}_{m} \right] \Bigr) - \theta^{2} \mathrm{Tr} \left[ B^{2} \rho \right].
\end{split}
\end{equation}

As mentioned in the main text, the DEC is applicable to the definition of the locally uniform QRMS disturbance \cite{Ozawa21}:
\begin{equation}
\bar{\eta}_{\text{O}}(B, \mathbf{M}, \rho) := \sup_{t \in \mathbb{R}} \eta_{\text{O}} (B, \mathbf{M},  e^{-i t B}\rho e^{i t B}).
\end{equation}
Namely, we can write this as
\begin{equation}
\begin{split}
\langle\bar{P}^{X}_{+}\rangle^{\text{Out}}_{\mathbf{P'}} &= \sum_{m} \mathrm{Tr} \Bigl[ ( I \otimes  P^{X}_{+}  ) V^{\dagger}(\theta) (M_{m}\otimes I) V(\theta) ( \bar{\rho} \otimes \proj{+} ) V^{\dagger}(\theta) (M_{m}^{\dagger}\otimes I) V(\theta) \Bigr] \\
&= 1 - \theta^2 \sum_{m} \mathrm{Tr} \left[ [M_{m},B] \bar{\rho} [M_{m},B]^{\dagger} \right] + O(\theta^{3})
\end{split}
\end{equation}
and
\begin{equation}
\begin{split}
\bar{\eta}_{\text{O}}(B, \mathbf{M}, \rho) = \sup_{t \in \mathbb{R}}\lim_{\theta \rightarrow 0} \frac{\langle\bar{P}^{X}_{+}\rangle^{\text{In}}_{\mathbf{P'}} - \langle\bar{P}^{X}_{+}\rangle^{\text{Out}}_{\mathbf{P'}}}{\theta^{2}},
\end{split}
\end{equation}
by setting $\bar{\rho} = e^{-i t B} \rho e^{i t B}$.

\subsection{B. Reason for interpreting measured system's QRMS disturbance as weak probe's decoherence}
The one of the key features of our approach is to introduce the weak interaction $V(\theta)$ between the measured system $\mathbf{S}$ and the weak probe system $\mathbf{P'}$ before the measurement $\{M_{m}\}$ to make a correlation in the composite system $\mathbf{S}+\mathbf{P'}$.
The other one is to associate the QRMS disturbance in the measured system $\mathbf{S}$ with a deviation between the expectation values of the projection observable $P^{X}_{+}$ for the input and output states in the weak probe system $\mathbf{P'}$.
The weak probe system $\mathbf{P'}$, as its name suggests, serves as a sensor to detect the QRMS disturbance.
By referencing the weak probe system $\mathbf{P'}$, we can evaluate the QRMS disturbance and achieve a clear operational mechanism as the evaluation method.
Since the input state of the weak probe system $\mathbf{P'}$ is the eigenstate $\ket{+}$ associated with the eigenvalue $+1$ of $X$, the deviation between the input and output states corresponds to the reduction in the off-diagonal elements of the output state represented by the density operator.
Therefore, by examining the difference in the expectation value of $P^{X}_{+}$ (or $X$) between the input and output states, we can measure the amount of reduction in the off-diagonal elements of the output state, which indicates the decoherence.
Through the weak interaction $V(\theta)$ and its inverse operation  $V^{\dagger}(\theta)$, we can propagate the observable correlation created by the measuring process onto the relative phase shift of the weak probe's state, making it detectable as the decoherence.
In this way, we can obtain the QRMS disturbance from the decoherence of the weak probe system $\mathbf{P'}$.

Indeed, the above statement can be rephrased in the context of the formulation, which quantifies a quantum coherence by $\ell_{1}$ norm.
Given a fixed reference basis (incoherent basis) $\{\ket{i}\}$, the measure of quantum coherence \cite{Baumgratz14} with respect to a state $\rho$ is defined by
\begin{equation}
\begin{split}
C_{\ell_{1}}(\rho):=\sum_{i\neq j}|\rho_{ij}|.
\end{split}
\end{equation}
If the state $\rho_{\text{inc}}$ is of the form $\rho_{\text{inc}}=\sum_{i}\rho_{i}\proj{i}$, $\rho_{\text{inc}}$ has no coherence (or $\rho_{\text{inc}}$ is incoherent) with regard to $\{\ket{i}\}$, i.e., $C_{\ell_{1}}(\rho_{\text{inc}})=0$.

Consider a completely positive trace-preserving (CPTP) map $\calD$, the measure of quantum decoherence by $\calD$ with respect to a input state $\rho^{\text{In}}$ is defined as
\begin{equation}
\begin{split}
D_{\ell_{1}}(\rho^{\text{In}},\calD):=C_{\ell_{1}}(\rho^{\text{In}})-C_{\ell_{1}}(\rho^{\text{Out}}),
\end{split}
\end{equation}
where $\rho^{\text{Out}}:=\calD(\rho^{\text{In}})$.

Denote the output state of $\mathbf{P'}$ from the disturbance evaluation process of the DEC as $\rho^{\text{Out}}_{\mathbf{P'}}:=\calD^{\text{DEC}}_{\mathbf{P'}}(\rho^{\text{In}}_{\mathbf{P'}})$, where $\calD^{\text{DEC}}_{\mathbf{P'}}(\cdot):=\sum_{m} \mathrm{Tr}_{\mathbf{S}} \Bigl[ V^{\dagger}(\theta) (M_{m}\otimes I) V(\theta) ( \rho \otimes  \cdot  ) V^{\dagger}(\theta) (M_{m}^{\dagger}\otimes I) V(\theta) \Bigr]$ and $\rho^{\text{In}}_{\mathbf{P'}}=\proj{+}$.
Then, the decoherence by $\calD^{\text{DEC}}_{\mathbf{P'}}$ with respect to the input state $\rho^{\text{In}}_{\mathbf{P'}}$ can be obtained as
\begin{equation}
\begin{split}
D_{\ell_{1}}(\rho^{\text{In}}_{\mathbf{P'}},\calD^{\text{DEC}}_{\mathbf{P'}})&=C_{\ell_{1}}(\rho^{\text{In}}_{\mathbf{P'}})-C_{\ell_{1}}(\rho^{\text{Out}}_{\mathbf{P'}})\\
&= 1 - \sum_{\substack{\alpha,\beta=0,1\\\alpha\neq\beta}}\bra{\alpha}\sum_{m} \mathrm{Tr}_{\mathbf{S}}\Bigl[(M_{m}\otimes I - i \theta [M_{m},B] \otimes Z + \frac{\theta^{2}}{2}[B,[M_{m},B]] \otimes I + O(\theta^{3})) \Bigr.\\
&\Bigl. \quad \mbox{}\times ( \rho \otimes \proj{+} ) (M^{\dagger}_{m}\otimes I - i \theta [M^{\dagger}_{m},B] \otimes Z + \frac{\theta^{2}}{2}[B,[M^{\dagger}_{m},B]] \otimes I + O(\theta^{3})) \Bigr]\ket{\beta} \\
&= 1 - \sum_{\substack{\alpha,\beta=0,1\\\alpha\neq\beta}}\bra{\alpha}\sum_{m} \mathrm{Tr}_{\mathbf{S}}\Bigl[( M_{m} \rho M^{\dagger}_{m} \otimes |+\rangle\langle+| - i \theta [M_{m},B] \rho M^{\dagger}_{m} \otimes |-\rangle\langle+| \Bigr.\\
&\Bigl. \quad \mbox{}- i \theta M_{m} \rho [M^{\dagger}_{m},B] \otimes |+\rangle\langle-| - \theta^2 [M_{m},B] \rho [M^{\dagger}_{m},B] \otimes |-\rangle\langle-| \Bigr.\\
&\Bigl. \quad \mbox{}+ \frac{\theta^{2}}{2} [B,[M_{m},B]] \rho M^{\dagger}_{m} \otimes |+\rangle\langle+| + \frac{\theta^{2}}{2} M_{m} \rho [B,[M^{\dagger}_{m},B]] \otimes |+\rangle\langle+| + O(\theta^{3})) \Bigr]\ket{\beta} \\
&= 1 - \sum_{m} \mathrm{Tr} \Bigl[ M_{m} \rho M^{\dagger}_{m} + \theta^2 [M_{m},B] \rho [M^{\dagger}_{m},B] + \frac{\theta^{2}}{2} [B,[M_{m},B]] \rho M^{\dagger}_{m} + \frac{\theta^{2}}{2} M_{m} \rho [B,[M^{\dagger}_{m},B]] \Bigr] + O(\theta^{3})\\
&= - 2 \theta^2 \sum_{m} \mathrm{Tr} \left[ [M_{m},B] \rho [M^{\dagger}_{m},B] \right] + O(\theta^{3})\\
&= 2 \theta^2 \sum_{m} \mathrm{Tr} \left[ [M_{m},B] \rho [M_{m},B]^{\dagger} \right] + O(\theta^{3}).
\end{split}
\end{equation}
By taking the derivative of $D_{\ell_{1}}(\rho^{\text{In}}_{\mathbf{P'}},\calD^{\text{DEC}}_{\mathbf{P'}})$ with respect to $\theta^{2}$, as in Eqs.~\eqref{s.eq:dist_p} and \eqref{s.eq:dist_x}, we obtain the QRMS disturbance as follows:
\begin{equation}
\begin{split}
\label{s.eq:dist_d}
\eta^{2}_{\text{O}}=\lim_{\theta\ra0}\frac{D_{\ell_{1}}(\rho^{\text{In}}_{\mathbf{P'}},\calD^{\text{DEC}}_{\mathbf{P'}})}{2\theta^{2}}.
\end{split}
\end{equation}
Eq.~\eqref{s.eq:dist_d} indicates the relations $D_{\ell_{1}}(\rho^{\text{In}}_{\mathbf{P'}},\calD^{\text{DEC}}_{\mathbf{P'}})=\langle X\rangle^{\text{In}}_{\mathbf{P'}}-\langle X\rangle^{\text{Out}}_{\mathbf{P'}}$ and $D_{\ell_{1}}(\rho^{\text{In}}_{\mathbf{P'}},\calD^{\text{DEC}}_{\mathbf{P'}})=2(\langle P^{X}_{+}\rangle^{\text{In}}_{\mathbf{P'}}-\langle P^{X}_{+}\rangle^{\text{Out}}_{\mathbf{P'}})$.

\section{\label{s.sec:sim_set_ec}\two. Simulation and experiment settings and evaluation criteria}
\subsection{A. Simulation and experiment details}
The theoretical value of the QRMS disturbance for each measurement on the input state $\rho = \proj{+i}$ used in our study is calculated below:
\begin{description}
   \item[Projective measurement of $X$] The measurement operators are $M^{X}_{\pm} := P^{X}_{\pm} = |\pm\rangle\langle\pm|$, and the commutation relation is $[|\pm\rangle\langle\pm|,X] = 0$.
Thus,
\begin{equation}
\eta^{2}_{\text{O}}(X) = \sum_{m = \pm} \mathrm{Tr} \left [M^{X}_{m},X] \proj{+i} [M^{X}_{m},X]^{\dagger} \right] = 0.
\end{equation}
\end{description}

\begin{description}
   \item[Projective measurement of $Y$] The measurement operators are $M^{Y}_{\pm} := P^{Y}_{\pm} = |\pm i \rangle\langle \pm i|$, and the commutation relation is
\begin{equation}
\begin{split}
[|\pm i \rangle\langle \pm i|,X] &= \frac{1}{2} [ (|0 \rangle\langle 0| + |1 \rangle\langle 1| \pm i |1 \rangle\langle 0| \mp i |0 \rangle\langle 1|) X \\
&\quad \mbox{}- X (|0 \rangle\langle 0| + |1 \rangle\langle 1| \pm i |1 \rangle\langle 0| \mp i |0 \rangle\langle 1|) ] \\
&= \frac{1}{2} [ (|0 \rangle\langle 1| + |1 \rangle\langle 0| \pm i |1 \rangle\langle 1| \mp i |0 \rangle\langle 0|) \\
&\quad \mbox{}- (|1 \rangle\langle 0| + |0 \rangle\langle 1| \pm i |0 \rangle\langle 0| \mp i |1 \rangle\langle 1|) ] \\
&= \frac{1}{2} (\mp 2 i |0 \rangle\langle 0| \pm 2 i |1 \rangle\langle 1|) = \mp i Z.
\end{split}
\end{equation}
Hence,
\begin{equation}
\begin{split}
\eta^{2}_{\text{O}}(X) &= \sum_{m = \pm} \mathrm{Tr} \left[ [M^{Y}_{m},X] |+i\rangle\langle+i| [M^{Y}_{m},X]^{\dagger} \right] = - \sum_{m = \pm } \mathrm{Tr} \left[ [M^{Y}_{m},X] |+i\rangle\langle+i| [M^{Y \dagger}_{m},X] \right] \\
&= - \mathrm{Tr} \left[ (-iZ) |+i\rangle\langle+i| (-iZ) \right] - \mathrm{Tr} \left[ (iZ) |+i\rangle\langle+i| (iZ) \right] = 2.
\end{split}
\end{equation}
\end{description}

\begin{description}
   \item[Projective measurement of $Z$] The measurement operators are $M^{Z}_{\pm 1} := P^{Z}_{\pm} = |0/1 \rangle\langle 0/1|$, and the commutation relation is
\begin{equation}
[|0/1 \rangle\langle 0/1|,X] = |0/1 \rangle\langle 0/1| X - X |0/1 \rangle\langle 0/1| = \mp i Y.
\end{equation}
As a result,
\begin{equation}
\begin{split}
\eta^{2}_{\text{O}}(X) = &\sum_{m = \pm} \mathrm{Tr} \left[ [M^{Z}_{m},X] |+i\rangle\langle+i| [M^{Z}_{m},X]^{\dagger} \right] = - \sum_{m = \pm } \mathrm{Tr} \left[ [M^{Z}_{m},X] |+i\rangle\langle+i| [M^{Z \dagger}_{m},X] \right] \\
&= - \mathrm{Tr} \left[ (-iY) |+i\rangle\langle+i| (-iY) \right] - \mathrm{Tr} \left[ (iY) |+i\rangle\langle+i| (iY) \right] = 2.
\end{split}
\end{equation}
\end{description}

In the quantum circuit implementations, the coupling strengths correspond to the magnitude of rotation angles in gate operations.
The weak measurement in the WMM consists of one $\mathsf{S}$ gate, one $\mathsf{R_{X}}$ gate, two $\mathsf{H}$ gates, and one $\mathsf{CNOT}$ gate [see Eq.~\eqref{eq:W} and Fig.~\ref{fig:dem}(b)].
Accordingly, the coupling strength of the weak interaction of the weak measurement in the WMM is adjusted by the rotation angle of the rotation gate $\mathsf{R_{X}}$ around the $X$-axis.
On the other hand, the weak interaction in the DEC consists of one $\mathsf{R_{Z}}$ gate, two $\mathsf{H}$ gates, and two $\mathsf{CNOT}$ gates [see Fig.~\ref{fig:dem}(c)].
The coupling strength of the weak interaction in the DEC is adjusted by the rotation angle of the rotation gate $\mathsf{R_{Z}}$ around the $Z$-axis.

\subsection{B. Reason for setting the evaluation criteria}
In our simulation and experiment, we have set strict criteria, as described in Methods, to evaluate the three methods, i.e. the TSM, WMM, and DEC.
This is because the performance of the three methods cannot be adequately evaluated solely on the basis of the bias and standard deviation.
Therefore, we have adopted the (classical) root-mean-square error $e$ as the comprehensive evaluation measure, which has the following relation:
\begin{equation}
\begin{split}
e^{2}-\sigma^2 &= \frac{1}{n} \sum_{j=1}^{n}\left[(a_j-a)^2-(a_j-m)^2\right] \\
&= \frac{1}{n} \sum_{j=1}^{n}\left[(a_j-a)+(a_j-m)\right]\left[(a_j-a)-(a_j-m)\right] \\
&= \frac{1}{n} \sum_{j=1}^{n}(2a_j-a-m)(m-a) \\
&= \frac{ m - a }{n} \sum_{j=1}^{n}(2a_j-a-m)\\
&= (m-a)(2m-a-m) \\
&=(m-a)^2 \\
&=b^{2}.
\end{split}
\end{equation}

\section{\label{s.sec:qem_sim}\three. Quantum error mitigation used in the experiments}
The quantum error mitigation (QEM) is a method to reduce the effect of noise from the outputs of circuits in quantum computing.
Here, we briefly explain the theory and implementation of the QEM methods, which are used in our experiments.

\subsection{A. Readout error mitigation}
In the readout process of the qubit measurement using the measuring device, two types of error occur: bit flip error with respect to classical labels and state change error that deviates from the ideal state reduction.
The goal of readout error mitigation (REM) is to compensate for the flipping of readout labels \cite{Chen19,Sarovar20,Bravyi21}.
REM achieves this by determining the confusion matrix (calibration matrix) that transforms the ideal output probability distribution from the noiseless detector to the noisy output probability distribution from the noisy detector; and then the ideal output probability distribution would be obtained by applying the (pseudo)inverse of the confusion matrix to the noisy output probability distribution.

\subsubsection{Theory of REM}
The key idea of REM is to estimate the confusion matrix and apply its (pseudo)inverse operation to the noisy output probability distribution.
Specifically, this is achieved through the following two steps.

\noindent
\textbf{1. Performing the quantum detector tomography:}

\noindent
Quantum detector tomography characterises the noisy detector by determining the confusion matrix $C$.
Denote the ideal output probability distribution by $p_{\text{ideal}}$, and the noisy output probability distribution is given by $p_{\text{noisy}}=C p_{\text{ideal}}$ using $C$.
Quantum detector tomography is realised by preparing the ground state, measuring one of the computational bases, and constructing $C$ from the obtained probabilities.

\noindent
\textbf{2. Applying the inverse confusion matrix:}

\noindent
Denote by $C_{\text{est}}$ the estimate of $C$ obtained from step 1.
Apply $C^{-1}_{\text{est}}$ to $p_{\text{noisy}}$ and obtain $p_{\text{est}}=C^{-1}_{\text{est}} p_{\text{noisy}}$.
In order to ensure that $p_{\text{est}}$ does not take negative probabilities or other non-physical values, the least-squares method is taken to obtain a pseudoinverse $C^{-1}_{\text{est}}$ using the Euclidean norm as $p_{\text{est}}^{*}=\arg\min_{p_{\text{est}}}(\|C p_{\text{est}}-p_{\text{noisy}}\|)$.

\subsection{B. Zero noise extrapolation}
Zero noise extrapolation (ZNE) is a straightforward error mitigation technique that can be applied with a relatively low sampling cost in many practical scenarios.
It is important to note that in certain cases, the extrapolated results may display a significant bias.
ZNE is not effective when the low-degree polynomial curve obtained by fitting the noisy expectation values fails to align with the zero-noise limit.
Nevertheless, one major advantage of ZNE is that it can be applied without requiring intricate knowledge of the underlying noise model, making it a powerful and versatile error mitigation technique.

\subsubsection{Theory of ZNE}
The fundamental concept of ZNE involves executing the circuit at various noise levels (error rates) and utilising the results to extrapolate the noiseless expectation value of an observable.
This process consists of two steps:

\noindent
\textbf{1. Applying the noise scaling:}

\noindent
There are two approaches that can be used to accomplish this: a continuous method at the noise level \cite{Temme17,Kandala19} and a digital method at the gate level \cite{Li17,Dumitrescu18}.
The former is typically represented by a technique called pulse stretching.
The latter is primarily implemented by unitary folding, identity insertion scaling, and Pauli twirling.
In this study, we adopt the unitary folding method \cite{Giurgica20} to intentionally increase the depth of the gates for boosting physical errors to effectively emphasise the influence of noise.
In unitary folding, we apply a mapping $U \mapsto U(U^{\dagger}U)$, which means that a sequence of $2n+1$ noisy unitary operations approximates one unitary operation with a physical error that is about $2n+1$ times greater.
The application of this mapping can be either global or local, depending on the situation.

\noindent
\textbf{2. Extrapolating to the noiseless limit:}

\noindent
In order to estimate the noise-free expectation value, there are various extrapolation methods (fitting a curve to the measured expectation values at various noise levels) available, including linear, polynomial, and exponential approaches.
In our study, we employ Richardson extrapolation, which assumes a valid Taylor expansion with a polynomial function of the error rates, considering higher-order terms to be negligible.
Let us consider a stochastic noise process labelled by $m$ in the following form:
\begin{equation}
\mathcal{E}_{m} := (1-\epsilon_{m})\mathcal{I}+\epsilon_{m}\mathcal{N}_{m},
\end{equation}
where $\epsilon_{m}$ is a noise parameter (small error rate).
Here, we use the superoperators; $\mathcal{I}$ is the identity map and $\mathcal{N}_{m}$ is the noisy map.
Suppose that we measure the expectation value of an observable $A$.
If we focus on one of the noise sources in realistic hardware, say $\mathcal{E}_{m}$, the expectation value $\langle A \rangle$ can be expanded in a Taylor series as a function of $\epsilon_{m}$:
\begin{equation}\label{eq:taylor_exp_a}
\langle A \rangle (\epsilon_{m}) = \langle A \rangle (0) + \sum^{n}_{k=1} A_{k} \epsilon^{k}_{m} + O(\epsilon^{n+1}_{m}).
\end{equation}
Note that in Eq.~\eqref{eq:taylor_exp_a}, $\langle A \rangle (0)$ represents the noiseless ideal expectation value and $A_{k}$ represents the coefficients obtained by Taylor-expanding $\langle A \rangle (\epsilon_{m})$ with respect to $\epsilon_{m}$ as a variable.
If we consider handling other sources of noise in addition to $\mathcal{E}_{m}$, we can perform a multivariable Taylor expansion of $\langle A \rangle$ with respect to each noise parameter.
Therefore, by introducing extrapolation for the expectation value of a single variable, the method can naturally be extended to multiple variables.
For simplicity, we consider only one noise parameter by  setting $\epsilon_{m}=\epsilon$ from now on.

To estimate the ideal measurement result $\langle A \rangle (0)$, we perform multiple measurements for $\langle A \rangle (\epsilon)$, with varying noise rates.
Although it may not be possible to decrease the error rate, we can effectively increase it using the aforementioned methods.
Hence, we consider multiple noisy measurements $\langle A \rangle (\alpha_{k}\epsilon)$, with different error rates $\alpha_{k}$, where $1<\alpha_{0}<\cdots<\alpha_{n}$.
We can then apply Eq.~\eqref{eq:taylor_exp_a} up to the $n$-th order and employ Richardson extrapolation to approximate $\langle A \rangle (0)$ as
\begin{equation}
\begin{split}
\langle A \rangle_{\text{est}} (0) &= \sum^{n}_{k=1} \beta_{k} \langle A \rangle (\alpha_{k}\epsilon), \\
&= \langle A \rangle (0) + O(\epsilon^{n+1}_{m}).
\end{split}
\end{equation}
Under the conditions of $\sum^{n}_{k=0}\beta_{k}=1$ and $\sum^{n}_{l=0}\beta_{l}\alpha^{k}_{l}=0$, the coefficients $\beta_{k}$ can be expressed as 
\begin{equation}
\beta_{k}=\prod_{i \ne k}\frac{\alpha_{i}}{\alpha_{k}-\alpha_{i}}.
\end{equation}
For details regarding the measurement cost and the variance in this method, refer to \cite{Endo21}.

\subsection{Implementation of REM and ZNE}
We generate a set of noise-scaled circuits by applying a unitary folding with different scale\_factor values for a quantum circuit defined via Qiskit.
To do this, we perform the mapping $U \mapsto U[(U^{\dagger}U)^{(\text{scale\_factor}-1)/2}]$ to a random local subset of individual gates of the input circuit.
The noise-scaled circuit is more sensitive to gate errors because it contains a number of gates approximately equal to r $g\times$scale\_factor, where $g$ is the number of gates in the input circuit, meaning that the number of gates is scaled exactly as dictated by the value of scale\_factor.
After that, we execute the noise-scaled circuits on the noisy backend to obtain a set of noise-scaled expectation values.
Here, scale\_factor is set to $\{ 1, 3, 5, 7, 9\}$, and we average over the five evaluations of each noise-scaled expectation value.
We store the outcomes (a set of noise-scaled expectation values) of the computations conducted at the selected noise level (the execution of the noise-scaled circuits) and estimate the zero-noise value by fitting the results using Richardson extrapolation.